\documentclass[12pt,preprint]{aastex}

\newcommand{\kms}{km~s$^{-1}$}

\begin{document}

\title{Star Formation and AGN in the Core of the Shapley Supercluster:\\
A VLA Survey of A3556, A3558, SC1327-312, SC1329-313, and A3562}

\author{Neal A. Miller\altaffilmark{1,2,3}} 
\email{nmiller@pha.jhu.edu}

\altaffiltext{1}{Jansky Fellow of the National Radio Astronomy Observatory. The National Radio Astronomy Observatory is a facility of the National Science Foundation operated under cooperative agreement by Associated Universities, Inc.}
\altaffiltext{2}{Department of Physics and Astronomy, Johns Hopkins University, 3400 N. Charles Street, Baltimore, MD 21218}
\altaffiltext{3}{Visiting Astronomer, CTIO, National Optical Astronomy Observatories, which is operated by the Association of Universities for Research in Astronomy, Inc., under cooperative agreement with the National Science Foundation.}

\begin{abstract} 
The core of the Shapley supercluster (A3556, A3558, SC1327-312, SC1329-313, and A3562) is an ideal region in which to study the effects of cluster mergers on the activity of individual galaxies. This paper presents the most comprehensive radio continuum investigation of the region, relying on a 63-pointing mosaic obtained with the Very Large Array yielding an areal coverage of nearly 7 square degrees. The mosaic provides a typical sensitivity of about 80 $\mu$Jy at a resolution of 16\arcsec, enabling detection of galaxies with star formation rates as low as 1 M$_\odot$ year$^{-1}$. The radio data are complemented by optical imaging in $B$ and $R$, producing a catalog of 210 radio-detected galaxies with $m_R \leq 17.36$ ($M_R \leq -19$). At least 104 of these radio-detected galaxies are members of the supercluster on the basis of public velocity measurements. Across the entire core of the supercluster, there appears to be a significant deficit of radio galaxies at intermediate optical magnitudes ($-21 \geq M_R > -22$). This deficit is offset somewhat by an increase in the frequency with which brighter galaxies ($M_R \leq -22$) host radio sources. More dramatic is the highly significant increase in the probability for fainter galaxies ($-20 \geq M_R > -21$) in the vicinity of A3562 and SC1329-313 to be associated with radio emission. The radio and optical data for these sources strongly suggest that these active galaxies are powered by star formation. In conjunction with recent X-ray analysis, this is interpreted as young starbursts related to the recent merger of SC1329-313 with A3562 and the rest of the supercluster.

\end{abstract}

\section{Introduction}

\subsection{Cluster Mergers and Their Effect on Member Galaxies}

Mergers between clusters are exciting events, involving around $10^{64}$ ergs of kinetic energy. On cluster-wide scales the effects of the dissipation of this energy are evident in the intracluster medium (ICM). X-ray observations of the ICM in presumed cluster mergers often reveal non-spherical morphologies, sometimes with temperature variations and shocks. Similarly, the presence of diffuse radio emission in the form of relics and halos seems to correlate with known cluster mergers. In such a dynamic environment, what is the effect on individual galaxies?

Any answers are likely complicated by numerous, and often competing, effects. Galaxies in groups which are being accreted by clusters may undergo bursts of star formation as time-varying tidal fields transfer gas to the galaxies' centers \citep{bekki1999}. Similar ``pre-processing'' of galaxies is found in simulations where galaxy-galaxy interactions and tidal perturbations are enhanced in groups and substructures around clusters \citep{gnedin2003}. On a larger scale, when clusters merge with each other they develop shocks in their intracluster gas and interactions between individual galaxies and these shocks may lead to starbursts \citep{roettiger1996}. In the case of individual galaxies already within clusters, the successive influence of numerous high-speed passages may lead to starbursts \citep[``harrassment,'' e.g.][]{moore1998}. Gas-rich galaxies entering a cluster would experience ram pressure through interaction with the intracluster gas \citep{gunn1972}, possibly compressing their molecular clouds and initiating bursts of star formation \citep{dressler1983}. However, ram pressure is also a prime candidate for suppressing star formation in clusters, as it can efficiently remove the gas supply of the entering galaxies, thereby extinguishing any star formation \citep[e.g.,][]{fujita1999,quilis2000}. As a result, the issue of whether or not a starburst ever occurs is an open one, with some authors finding evidence for such starburst histories \citep[e.g.,][]{poggianti1999} with others favoring ``strangulation'' of star formation \citep[e.g.,][]{balogh1999}. Lastly, one should bear in mind that all of these arguments may be moot: the member galaxies of any single cluster may already be quiescent and gas poor, leaving little material available for transformation in a cluster merger.

Recently, radio investigations have been used to study galaxy activity in cluster mergers. Radio emission is an excellent indicator of galaxy activity, as it can result from active galactic nuclei (AGN) or star formation \citep[see ][for a review]{condon1992}. In fact, in the absence of any contamination due to an AGN the radio luminosity of a galaxy (typically measured at 1.4~GHz) is directly proportional to its star formation rate \citep[e.g.,][]{yun2001}. Much of the interest in radio investigations of cluster mergers was inspired by A2125, a merging cluster studied in tandem with the relaxed cluster A2645 in \citet{dwar1999} and \citet{owen1999}. The clusters are of the same richness and redshift, yet A2125 has about an order of magnitude more radio galaxies. Subsequent radio studies of presumed cluster mergers include the Shapley supercluster \citep[A3556, A3558, and A3562,][]{venturi1997, venturi2000, giacintucci2004}, A2255 \citep{my2255}, A2256 \citep{my2256}, a large sample of intermediate-redshift clusters \citep{morrison2003}, and a much more comprehensive analysis of A2125 \citep{owen2005a,owen2005b}. The resulting picture may not be so simple, with the fraction of galaxies hosting radio sources increased in some mergers (A2125, A2255), normal in others (A2256), and potentially even decreased in others (A3558). These differences might be explained by noting the examined populations and considering the timing of the merger. The former point is merely that the effects of a cluster merger are likely different on star-forming galaxies than they would be on brighter and more massive AGN, whereas the latter notes that cluster mergers proceed on $\sim$Gyr timescales while the radio emission is associated with events whose duration is about an order of magnitude less. Consequently, there are two clear directives for further observations: 1) to investigate previously-studied clusters to consistent and deep sensitivity limits, and 2) to investigate a larger number of merging clusters in order to better understand the role of merger timing.

\subsection{An Ideal Case Study: The Shapley Supercluster}

The Shapley supercluster represents an ideal testbed for further investigation of the effects of cluster mergers on the evolution of member galaxies. The true extent of the supercluster is not determined, as redshift surveys covering increasing areas of the sky have continued to reveal its enormous scale through interconnected structures \citep[e.g.,][]{drinkwater2004}. At the center of the supercluster lies A3558 (or Shapley 8), a very rich cluster (Abell richness class 4) flanked by A3556 to the west and A3562 as well as a pair of poor clusters identified in both the optical and X-ray \citep[SC1327-312 and SC1329-313, see e.g.,][]{breen1994,bardelli1998b} to the east. This collection of clusters forms the core of the supercluster, and has been studied extensively at a variety of wavelengths. Furthermore, all of these clusters lie in the plane of the sky at $z\sim0.048$ making this a particularly attractive supercluster for a single observational campaign. 

Several research collaborations have committed large amounts of optical observing time to the collection of hundreds of galaxy spectra in the Shapley supercluster, and the resulting dynamical assessments depict an active region rich in substructure. The supercluster has been studied on large scales (tens of Mpc) in an attempt to estimate its total mass \citep{quintana1995,quintana2000,reisenegger2000,drinkwater2004} since it lies in the direction of the Local Group's motion relative to the cosmic microwave background. \citet{bardelli1998b} presented a substructure analysis of the core of the supercluster, drawing on the velocities presented in a sequence of works \citep{bardelli1994,bardelli1996,bardelli1998a}. They applied the {\scshape dedica} algorithm to identify up to 21 significant substructures within the supercluster core, and concluded that either of two hypotheses held: 1) the core of the Shapley supercluster represents a cluster-cluster merger viewed shortly after one cluster core has passed the other for the first time, or 2) the entire region is a complex of lesser mergers (such as groups merging with each other and with larger clusters).

Similarly, X-ray observations indicate a dynamically rich system. The cluster known as SC1327-312 was first identified on the basis of {\sl Einstein} data \citep{breen1994}, and subsequent studies with various X-ray observatories have continued to reveal interesting properties: for example, {\sl ROSAT} \citep{kull1999,ettori2000}, {\sl ASCA} \citep{hanami1999}, {\sl Beppo-SAX} \citep{bardelli2002}, and {\sl XMM-Newton} \citep{finoguenov2004}. All of the five clusters are linked by filamentary X-ray structure \citep{kull1999}, and appear to be interacting. \citet{hanami1999} suggested a dynamical sequence for the clusters in the chain including a relaxed poor cluster (SC1327-312), two rich ``developed'' clusters with substructure (A3558 and A3562), an ongoing merger (SC1329-313), and a candidate post-merger (A3556). Their interpretations were largely based on departures from expected relationships between $L_X$, $T$, and $\sigma_v$ (optical velocity dispersion). The ongoing merger in SC1329-313 was deduced from its fitted Fe K$\alpha$ line, indicating the cluster was out of ionization-equilibrium (the fitted energy of the iron line was consistent with the H-like iron K$\alpha$ line instead of the usual He-like line expected for the measured temperature of the cluster). \citet{bardelli2002} were only able to confirm this result for SC1329-313 at the 2$\sigma$ level, and instead argued that the whole system represented the aftermath of a major cluster-cluster merger seen after the first core passage, an interpretation also based on their optical and radio investigations. In this interpretation, the clusters other than A3558 are the remains of the merger. \citet{finoguenov2004} argued that SC1329-313 had recently passed through the northern outskirts of A3562. Despite the varied interpretations of the merger history and partners, it is clear that the core of the Shapley supercluster is a dynamically-rich system.

Prior radio works targeting specific regions within the system also reveal a complex environment. \citet{venturi1997} studied A3556 using the Australia Telescope Compact Array (ATCA) and Molonglo Observatory Synthesis Telescope (MOST), identifying nine cluster radio galaxies including the unusual narrow-angle tail source J1324-3138 \citep{venturi1998}. This appears to be the remnant of a radio galaxy, where the central engine has switched off possibly in response to the cluster merger event. \citet{venturi2000} extended the same general radio survey of the Shapley supercluster to include the cores of the other clusters in the chain, increasing the number of cluster radio galaxies to 28. Using these identifications to construct the radio luminosity function (RLF), they found a deficit of powerful radio galaxies in the system relative to the cluster RLF determined in \citet{ledlow1996}. They argued that cluster mergers may switch off existing radio sources, as appeared to be the case for J1324-3138. \citet{giacintucci2004} continued the overall survey, in particular covering the region around A3562. They identified 33 cluster radio galaxies (26 of which were not in the preceding works), and found that the deficit of powerful radio galaxies noted in \citet{venturi2000} was restricted to A3558. Among these sources were a number of lower radio luminosity sources, presumed to be starbursts.

In this paper, a comprehensive radio and optical view of the entire core of the supercluster is presented. The radio data were obtained with the NRAO Very Large Array (VLA) and consist of a 63-pointing mosaic which covers nearly 7 square degrees at a fairly uniform sensitivity of $\sim$80 $\mu$Jy per 16\arcsec{} beam. This makes it unique in that it covers the entire core of the Shapley supercluster in a single observational campaign, and implies that AGN and galaxies forming stars at rates as low as 1.0 M$_\odot$ year$^{-1}$ are detected. The radio data are complemented by new optical imaging in $B$ and $R_c$. The areal coverage and uniform sensitivity of these data represent the key advantages of this study: populations of active galaxies down to low activity levels may be studied across the full range of environmental conditions within the supercluster. This enables stronger conclusions about the potential correlation of increased activity within individual galaxies to regions of dynamical activity within the clusters which make up the supercluster.

For the sake of consistency with prior papers investigating the radio galaxy populations of nearby clusters, a cosmology with $H_0 = 75$ \kms{} Mpc$^{-1}$ and $q_0 = 0.1$ has been adopted. This yields a luminosity distance of 186.8 Mpc to the supercluster (using $z=0.048$), meaning $1^{\prime\prime} = 0.82$ kpc. For reference, these values become $D_L = 210.3$ Mpc and $1^{\prime\prime} = 0.93$ kpc for the {\it WMAP} cosmology with $H_0 = 71$ \kms{} Mpc$^{-1}$ and $\Omega_M = 0.27$, $\Omega_\Lambda = 0.73$.

\section{Data and Reductions}

\subsection{Radio}

The initial radio observations were performed in 2001 June with the VLA in its CnB configuration. This configuration, observing at a frequency of 1.4~GHz, is well suited to the study of star-forming galaxies in the Shapley supercluster with an angular resolution of $\sim$15\arcsec{} (12.4 kpc), meaning low surface brightness emission spread over the disk of a galaxy will not be missed. The extended north arm of the configuration also results in a more circular beam. The program was scheduled over five days, each consisting of a 6-hour block centered on transit of A3558 (see Table \ref{tbl-radobs} for a listing of all pointings and observation dates). Unfortunately, data were lost due to thunderstorms and strong winds stowing the array on 2001 June 20 and mechanical problems on 2001 June 21. To replace these data, a single additional 6-hour track was scheduled during the subsequent CnB configuration of the VLA in 2002 September. The conditions for these observations were less favorable, as they were performed during daytime where the Sun is a source of additional noise.

The observational strategy consisted of observing the core of the supercluster through a mosaic of individual VLA pointings. These were arranged in a hexagonal grid to provide nearly uniform sensitivity across the entire area, with an 18\arcmin{} spacing between adjacent pointings. This mosaic strategy is the same as that employed by the NRAO VLA Sky Survey \citep[NVSS,][]{condon98} and Faint Images of the Radio Sky at Twenty centimeters \citep[FIRST,][]{becker95} although with a tighter grid spacing. Originally, 70 pointings were planned but this was pared down to 63 due to the lost observing time. The seven pointings removed from the grid were taken from the edges to minimize the effect on the final mosaic coverage. The final 63 pointings are summarized in Table \ref{tbl-radobs}, and a footprint of the covered area may be viewed in Figure \ref{fig-foot}.

The radio data were obtained in line mode at 1.4~GHz, consisting of 28 channels each of 3.125~MHz bandwidth (two sets of seven channels at each polarization). Each pointing center was visited only once to minimize the amount of time spent moving among pointings. To account for the varying system temperature of the VLA at 1.4~GHz as a function of elevation, the dwell times were adjusted in order to achieve a similar sensitivity at each pointing. The minimum dwell time (for sources observed near transit) was 20 minutes, while sources at the beginning and ends of the runs were observed for 27 minutes. Although the author sure thought he was clever in doing this, he failed to respect the importance of good ($u,v$) coverage. For the most part, the pointings were prioritized such that the edges of the supercluster corresponded to those with poorer ($u,v$) coverage. To improve the ($u,v$) coverage of a few central pointings, they were revisited for brief periods ($\sim10$ minutes) during the 29 September 2002 observations. The effect of the ($u,v$) coverage will be addressed in more detail below. Flux calibration was achieved using 3C286, with calibration of phases and bandpasses performed from roughly hourly observations of the nearby calibrator source J1316-336. 

The NRAO's Astronomical Image Processing System (AIPS) was used to calibrate and image the data, following the usual procedures for 1.4~GHz data.\footnote{See {\url http://www.vla.nrao.edu/astro/guides/lowfreq/analyses/}} Briefly, the data for each pointing were calibrated (including antenna-specific weights) and reduced individually. Ideally, one would handle the ($u,v$) data for each pointing in a manner which both produces a consistent beam across all pointings and does so with good characteristics (i.e., a nice Gaussian with minimal sidelobes). This is where the effect of moving to each position in the pointing grid only once is most relevant. For several of the pointings at the edges of the pointing grid, the resulting beam is highly elongated. This renders it virtually impossible to obtain a consistent beam across all pointings, necessitating the compromise of choosing parameters which produce beams with the same area and using a circular restoring beam of equal area in the final maps. Table \ref{tbl-radobs} indicates the initial beam sizes. Each pointing was imaged in $\sim10$ fields, with four fields covering the primary beam and additional fields dedicated to outlying bright sources. Iterations of imaging and self-calibration were performed until the final images for each pointing were created. At this point, the cleaned sources were restored with a 16\arcsec{} circular beam for combining into the final mosaic. The rms noise of these individual pointings ranged from about 75 to 125 $\mu$Jy beam$^{-1}$.

The final mosaic was created by combining the reduced images for all 63 pointings. At this stage, the flanking fields dedicated to bright sources were ignored and only the four fields covering the primary beam for each pointing were used. The contribution of each image to the mosaic was properly weighted by its signal-to-noise, as discussed in \citet{condon98}, with the images truncated at the point where the response of the VLA had dropped to $30\%$ of its peak (i.e., each pointing images about a 40\arcmin{} diameter). The noise characteristics of the final mosaic are quite good, with a typical rms of about 80 $\mu$Jy beam$^{-1}$ over the area described as within 2 Mpc of the centers of the individual Abell clusters. The noisiest region within this area has a local noise around 115 $\mu$Jy beam$^{-1}$ and is found to the East of the center of Abell 3558. It is caused by several bright sources, particularly the quasar at J133019.1-312259 \citep[e.g.,][]{hewitt1993}. The cleanest regions have rms noise as low as 62 $\mu$Jy beam$^{-1}$. The final mosaic is available upon request from the author.

The catalog of radio sources was then created using the task SAD, which identifies peaks in a map and fits them with Gaussians. Parameters of importance are saved from these fits, including the source position, peak flux density, integral flux density, major and minor axis, plus position angle. SAD also creates a residual map, which was inspected to identify any sources for which Gaussian fits failed (usually strong extended sources). These were manually added to the radio source catalog. The relative flux scale was found to be consistent with the NVSS via comparison of the fluxes of unresolved sources.

\subsection{Optical}

Optical images in both $B$ and $R_c$ (hereafter referred to simply as ``$R$'') filters were obtained in 2002 March and April using the Cerro Tololo Inter-American Observatory 1.5m telescope with the Cassegrain Focus CCD Imager. Using the $f/7.5$ focus provided a wide field of view of 14\farcm8 with 0\farcs44 pixels. Even so, the large area of the core of the supercluster required numerous individual telescope pointings: 152 fields in each $B$ and $R$, arranged in a simple grid with a 14\arcmin{} spacing. Figure \ref{fig-foot} depicts the coverage of the optical fields in relation to the radio mosaic. Data for this project were collected on the nights of March 24 and 25 (kindly provided by Michael Ledlow during a run for a related program), and March 29, March 31, and April 1. Sky brightness was an issue, as these dates straddled the full Moon. The exposures were prioritized such that the $B$ images were collected during the darker conditions, and exposure times were adjusted slightly for the conditions on a nightly basis. The $R$ exposures ranged from 120 to 150 seconds, while the $B$ exposures were either 180 or 240 seconds. The seeing varied from 1\farcs1 (about the best possible with the $f/7.5$ focus) to 2\farcs0, with most exposures falling in the $1\farcs3 - 1\farcs4$ range.

Data reduction followed the standard reduction steps. The images were bias subtracted and then flat fielded using twilight frames collected during several of the nights. The astrometry was registered using about 20 unsaturated USNO A2.0 stars per field \citep{monet2000}, yielding rms errors of under 0\farcs25.  The photometry was set by observations of ``Selected Area'' fields from \citet{landolt1992}, performed nightly and at a range of airmass. The derived relationships were:

\begin{equation}
R = 23.29 - 2.5\log (cts) + 2.5 \log (t) - 0.09X
\end{equation}
\begin{equation}
B = 23.16 - 2.5\log (cts) + 2.5 \log (t) - 0.29X - 0.05(B - R)
\end{equation}
where $cts$ is the background-subtracted counts from the source, $t$ is the integration time measured in seconds, and $X$ is the airmass. The consistency of the photometry was checked using stars at the edges of the science exposures, where the overlap of the pointings produced numerous comparison objects (see below).

Lists of optical sources were generated by field using SExtractor \citep{bertin1996}. The most important parameters proved to be those related to the background as scattered light was an issue, particularly for the $R$ images collected during times of high sky brightness. This produced images with a gradient in the apparent background which often changed sharply along one edge of the CCD. Choosing a background mesh size too large failed to properly respond to this effect, while smaller background mesh sizes performed poorly around the larger real objects. In this latter case, some of the counts corresponding to the bigger elliptical galaxies and brighter saturated stars get included in the background resulting in extracted magnitudes for these objects which are fainter than their true magnitudes. Since one of the main uses of the optical magnitudes is to select galaxies brighter than prescribed limits, the chosen mesh size erred more toward correct handling of the scattered light. This means that the reported magnitudes for the brightest elliptical galaxies ($m_R \lesssim 14.5$) are up to $\sim0.1$ mags fainter than their real values. The derived magnitudes for all fields correspond to the fixed aperture size of 15\farcs9 in radius, which is the Gunn-Oke aperture \citep{gunn1975} at the assumed redshift of the supercluster ($z=0.048$). Initially, the color term in the above photometric equations was ignored and later introduced when the output catalogs were merged. Further, when the source catalogs for the individual fields were merged the magnitudes were also corrected for Galactic extinction using the values of \citet{schlegel1998}, based on the $A_B$ and $A_R$ values for the center of each 14\farcm8 field. Limiting magnitudes, based on inspection of number count histograms, ranged from 18.5 to 19.5 in $R$ and from 19.5 to 21 in $B$.

The consistency of the photometry from field to field was then checked using the regions of overlap. The magnitude errors reported by SExtractor are based simply on counting statistics, so this step essentially determines the errors resulting from application of the above photometric equations to the collected data. It would also reveal any magnitude zero point shifts as would occur under non-photometric conditions. All designated stars whose photometry was not flagged by SExtractor were used. The consistency of the $B$ photometry was excellent, with a standard deviation under 0.05 mag for objects with magnitudes brighter than $B=17$ and still under 0.15 mag for galaxies as faint as $B=18$. As previously suggested, the $R$ photometry was less consistent with a standard deviation of less than 0.1 mag out to $R=16$ and rising to 0.2 mag at $R=17$. For simplicity, these findings have been generalized and an additional error of 0.05 mag in $B$ and 0.1 mag in $R$ have been included in the reported photometry. Lastly, the merged catalog of all optical sources was edited to remove duplicates (see additional discussion below). 

\subsection{Source Identification}\label{sec-id}

The radio and optical source catalogs were then merged to create the final list of radio galaxies. The adopted conventions for creation of the radio galaxy catalog were:
\begin{itemize}
\item{The radio peak flux must be greater than 330 $\mu$Jy, chosen as a 5$\sigma$ detection in the regions of the mosaic map with the lowest noise.}
\item{Either the radio peak flux or integral flux must be greater than five times the local rms noise, as evaluated in a 6\arcmin{} box centered on the peak in the radio emission.}
\item{The galaxy must be brighter than $m_R = 17.36$, which corresponds to $M_R = -19$ in the adopted cosmology. This limit is one magnitude fainter than that used in \citet{my2255} and related papers, allowing the present study to sample cluster radio galaxies down to fainter levels. It is also safely brighter than the completeness limit of the worst $R$ band fields. Although radio sources with detected optical counterparts fainter than 17.36 are present in the data, most such radio galaxies are safely assumed background objects.}
\item{The separation between radio and optical position must be less than 7\arcsec{} (except in cases of extended powerful radio galaxies). The probability of chance coincidence of radio and optical objects satisfying the above criteria was evaluated by shifting the optical catalog by arbitrary amounts and re-performing the correlation. From this analysis, it is expected that about 4 sources in the final list are false. However, the false detection probability for sources within 2 Mpc of the cluster centers and with $M_R \leq -20$ and $L_{1.4} \geq 6.8 \times 10^{21}$ W Hz$^{-1}$ is much lower. These are the limits used for the comparison sample in the radio galaxy fraction analysis (see Section \ref{sec-rgfrac}), and the false detection probability corresponding to the 7\arcsec{} separation limit approximately matches that used in construction of that comparison sample.}
\end{itemize}

Perhaps the most critical step in source identification was a thorough visual inspection. The radio contours were overlaid on the optical images along with markers identifying sources in each the radio and optical catalogs. In the case of the radio data, this step allowed for extended radio galaxies to be identified in addition to any other problem cases not fit by SAD. The extended radio sources are often missed because the peak in their emission does not necessarily coincide with the optical location of the galaxy. For the optical data, the visual inspection provided a check on the SExtractor results including the star/galaxy segregation for problem objects such as close pairs of faint stars and galaxies. These false bright objects were removed from the catalog of optical sources. The use of markers for each of the radio and optical catalogs also enabled the removal of any duplications which were missed in the previous steps. Finally, a small number of optical galaxies in close proximity to bright stars were absent from the SExtractor catalogs. Magnitudes for these galaxies were determined manually by summing counts within irregular shaped apertures designed to avoid contamination by the nearby stars.

The final list is presented in Table \ref{tbl-radgals}. This table includes the optical position, magnitudes and color, integral radio flux, peak radio flux, a marker indicating whether the radio emission was unresolved, local noise at the location of the radio source, and separation between the optical and radio positions. Those sources for which the AIPS task JMFIT found a minimum size of zero for the major axis were assumed to be unresolved. For such sources, the fitted peak flux density is a better representation of the true flux of the source than the fitted integral radio flux \citep[e.g., see][]{owen2005a}. In addition, the NASA/IPAC Extragalactic Database\footnote{NED is operated by the Jet Propulsion Laboratory, California Institute of Technology, under contract with the National Aeronautics and Space Administration.} (NED) was searched to determine whether velocity information was available for each of the 210 entries in Table \ref{tbl-radgals}. A total of 123 galaxies had public velocity information, including at least 104 members of the Shapley supercluster (see next Section). This information is also presented in Table \ref{tbl-radgals}. In all subsequent text, the 210 radio-detected galaxies will be referred to simply as ``radio galaxies,'' with the 104 spectroscopically-confirmed members of the Shapley supercluster called ``cluster radio galaxies'' or sometimes ``cluster members.''

\section{Analysis}

\subsection{Overview of Radio Galaxy Population}\label{sec-overview}

To determine which radio galaxies belong to the Shapley supercluster, the individual cluster recession velocities and dispersions ($\sigma_v$) of \citet{bardelli1998a} were used. Allowing galaxies within $\pm 3\sigma_v$ of any cluster/substructure, the minimum and maximum velocities describing the supercluster are 10934 \kms{} and 17684 \kms. In practice, these limits correspond to those for the brighter galaxies in A3558 \citep[$<v> = 14309$ \kms{} and $\sigma_v = 1125$ \kms;][]{bardelli1998a}. As noted previously, there are 104 cluster member radio galaxies identified in the present study, nearly doubling the total identified in prior studies. These include all of the cluster radio galaxies reported in the prior radio studies of the supercluster \citep{venturi1997,venturi2000,giacintucci2004}, with the exception of two spirals with low integral flux.\footnote{13:32:05.6 -31:52:30 at 0.49 mJy and 13:30:52.1 -32:08:56 at 0.73 mJy \citep{giacintucci2004}. The local noises at these positions in the mosaic radio map are each about 75 $\mu$Jy beam$^{-1}$ rms.} Of the 17 radio galaxies with velocities formally placing them outside the supercluster, 6 are background radio sources and 11 are foreground. The majority of these foreground objects are likely related to the supercluster, having velocities greater than 9000 \kms{} and typically located in the eastern region of the studied area. The other three foreground galaxies belong to the Hydra-Centaurus wall \citep[$\sim4000$ \kms, see][]{drinkwater2004}. A further two radio galaxies were associated with galaxy pairs in NED. Although the galaxies hosting the radio emission did not have NED velocities, their companions did have cluster velocities and hence they may tentatively be considered cluster members which would bring the total to 106. There are 87 remaining radio galaxies for which no velocity data are available. Of these, 19 are probable background galaxies on the basis of their $B - R$ colors (see below).

Table \ref{tbl-radgals} includes nine galaxies with radio emission extended beyond their optical sizes, seven of which are cluster members (see Figures \ref{fig-132145} - \ref{fig-133542}). Of the cluster members, three have $L_{1.4} \geq 10^{23}$ W Hz$^{-1}$ and are thus in the realm of classical Fanaroff-Riley objects \citep{fr1974}: J132357-313845, J132802-314521, and J133331-314058. In each case, the radio morphology appears to be ``head-tail'' in nature. J132357-313845, depicted in Figure \ref{fig-132357}, is discussed at length in \citet{venturi1998}. Similarly, J133331-314058 (Figure \ref{fig-133331}) is analyzed in detail in \citet{venturi2003}. The third possible head-tail source, J132802-314521, consists of a strong core centered on the galaxy plus what appears to be a diffuse tail (Figure \ref{fig-132802}). It is possible that these features are unrelated and the potential tail is a background source. However, the luminosity of the core portion of the radio emission is $1.1 \times 10^{23}$ W Hz$^{-1}$. Such high luminosities are usually associated with extended emission, lending credence to the association of the diffuse emission as a low surface brightness extension of the source. With this emission, the luminosity becomes $1.3 \times 10^{23}$ W Hz$^{-1}$. Similarly, the optical magnitude of the galaxy ($M_R = -22.1$) is typical of Fanaroff-Riley sources. A caveat to this interpretation is that the higher resolution observations of \citet{venturi2000} (10\arcsec $\times$ 5\arcsec) did not resolve the core emission, as might be expected if it were a jet associated with the diffuse emission. The detection of up to three head-tail sources across the clusters at the core of the Shapley supercluster is typical of rich clusters in general \citep[e.g., see][]{ledlow1995}. An additional four extended radio galaxies belonging to the cluster have general morphologies which would place them among Fanarof-Riley objects, although in each case the luminosity is below $10^{23}$ W Hz$^{-1}$. These include: J132145-310300 (Figure \ref{fig-132145}), with emission extended to the north and south suggestive of weak ``jets'' but $L_{1.4}$ of only $3.2 \times 10^{22}$ W Hz$^{-1}$; J132206-314616 \citep[Figure \ref{fig-132206} and see][]{venturi1997}, which has the appearance of a compact double but a luminosity of only $3.0 \times 10^{22}$ W Hz$^{-1}$, much lower than the $L_{1.4} \geq 10^{25}$ W Hz$^{-1}$ typical of such sources; J133048-314325 (Figure \ref{fig-133048}), with $L_{1.4} = 3.1 \times 10^{22}$ W Hz$^{-1}$ and discussed in more detail in Section \ref{ssec-evolution}; and J133542-315354 (Figure \ref{fig-133542}) at $6.0 \times 10^{22}$ W Hz$^{-1}$, which might be contaminated by emission from faint background galaxies to the west. The remaining two extended radio galaxies are strong background AGN: J132210-312805 (Figure \ref{fig-132210}) exhibits the morphology of a strong double-lobed radio source, and J132950-312258 (Figure \ref{fig-132950}) with a NED velocity of 58775 \kms{} implying $L_{1.4} = 9.7 \times 10^{23}$ W Hz$^{-1}$.

\subsection{The Cluster Red Sequence and Radio Galaxy Colors}

An advantage to having optical data in two filters is that the cluster red sequence (also referred to as the E/S0 ridge line) may be identified. This is useful in characterizing the radio galaxies, particularly those for which spectra are not available. Galaxies which lie along the red sequence are most likely supercluster member ellipticals, while galaxies above the red sequence (i.e., larger $B - R$ for a given $R$ magnitude) may safely be assumed to be background objects. Bluer objects may be star-forming galaxies associated with the supercluster, but may also be foreground or background objects. In most cases, NED velocities enable this determination.

The red sequence was fit using the method of \citet{lopezcruz2004}, with a few small variations. Briefly, the red sequence is parametrized as a line such that $y_i = a + bx_i$ where $y_i$ is the $B - R$ color for a given galaxy, $x_i$ is its $R$ magnitude, and $a$ and $b$ are the intercept and slope, respectively. To fit the red sequence, the deviation of each galaxy from a specified $a$ and $b$ is determined. The distribution of these deviations is evaluated using the biweight location and scale \citep{beers1990}, and a range of $a$ and $b$ are searched to find the pair that minimizes these quantities. All galaxies brighter than $m_R=17.36$ were used in the fits. Instead of using a radial cutoff of 0.2$r_{200}$ as in \citet{lopezcruz2004}, in the present paper the red sequence was fit for galaxies within 2 Mpc of the adopted cluster centers. In addition, the galaxies used in fitting the red sequence were iteratively 3$\sigma$ clipped to remove outliers. This clipping generally only removed very red background galaxies. The resulting fits are depicted in Figure \ref{fig-redseq} and summarized in Table \ref{tbl-rsfit}.

The fits for the individual clusters have steeper slopes than those of similar redshift clusters reported in \citet{lopezcruz2004}. The relationship derived in that study predicts a slope of $-0.047$, shallower than all of the fitted slopes in the present study. However, this difference is unlikely to be significant. A simple test is to evaluate the colors predicted by the different fits in Table \ref{tbl-rsfit} for representative $R$ magnitudes of cluster galaxies. For galaxies with $m_R$ from 12.7 to 17.3, the fits for the individual clusters produce colors which vary by only 0.1 -- less than the photometric errors for the galaxies. For context, the derived relation for the full galaxy sample regardless of radial separation from the cluster cores has a slope of $-0.054$, comparable to the generalized results of \citet{lopezcruz2004}. As the primary purpose of the red sequence fits for this paper is the identification of background galaxies and categorization of cluster radio galaxies as star-forming or AGN, further analysis of the red sequence fits is deferred.

Colors for the radio galaxies presented in Table \ref{tbl-radgals} are depicted in Figure \ref{fig-colors}. The change in the nature of the galaxy activity is apparent. The optically-brightest galaxies predominantly lie on the cluster red sequence and are presumably powered by AGN. Moving to fainter optical magnitudes, bluer colors are more prevalent as star formation becomes the origin for the radio emission.

\subsection{Radio Galaxy Fractions}\label{sec-rgfrac}

\subsubsection{Analysis by Cluster}
One of the key science drivers for studying the Shapley supercluster is to assess the fraction of active galaxies in relation to other clusters of galaxies. For this purpose, the analysis presented in \citet[][hereafter MO03]{my2255} for A2255 is performed for the individual clusters in the Shapley supercluster using the 18 nearby clusters presented in \citet{miller2001} as a comparison sample. Briefly, the radio galaxy fraction for a given cluster is evaluated as the number of radio galaxies with luminosity greater than a prescribed limit of $6.8 \times 10^{21}$ W Hz$^{-1}$ divided by the total number of galaxies. This radio luminosity is based on the completeness limit of the NVSS at the most distant cluster of the comparison sample, and assumes a spectral index of 0.7 for the k correction (where $S_\nu \propto \nu ^ {-\alpha}$). In the Shapley supercluster, it corresponds to a flux of 1.64 mJy which is well above the lower limits of the full radio mosaic. Velocity information is applied in order to remove foreground and background objects from the radio galaxies which comprise the numerator of the fraction. The allowed velocity ranges for each cluster, taken from \citet{bardelli1998a} and corresponding to $\pm 3\sigma_v$ about each cluster's systemic velocity, were: A3556, 12428 -- 16286 \kms; A3558, 10934 -- 17391 \kms; A3562, 11753 -- 17231 \kms; SC1327-312, 12771 -- 16971 \kms; and SC1329-313, 12520 -- 15921 \kms. Complete velocity information is unavailable for the denominator, so the galaxy counts are corrected for contamination as in MO03 by assuming $N = \mathcal{N}10^{0.6m}$, where $\mathcal{N} = 1.26 \times 10^{-5}$ galaxies steradian$^{-1}$. This value of $\mathcal{N}$ was derived directly from regions outside the comparison sample clusters. The analysis is performed for galaxies with $M_R \leq -20$ and within 2 Mpc projected separation of the respective cluster centers. The adopted center positions were (J2000 coordinates): A3556, 13:24:06.2 -31:39:38; A3558, 13:27:54.8 -31:29:32; A3562, 13:33:31.8 -31:40:23; SC1327-312, 13:29:47.0 -31:36:29; and SC1329-313, 13:31:36.0 -31:48:46. As in MO03, the radio galaxy fractions are evaluated for optically bright objects ($M_R \leq -22$), intermediate objects ($-22 < M_R \leq -21$, a range which includes $M^*$), and faint objects ($-21 < M_R \leq -20$). The actual testing is performed using a $\chi ^2$ statistic which evaluates the probability that a given cluster deviates from the fraction predicted by the pooled comparison sample.

Table \ref{tbl-abs} is provided to assist the reader in identifying which galaxies from Table \ref{tbl-radgals} enter the radio galaxy fraction analysis. In addition to the galaxies' optical positions, it includes their absolute $R$ magnitudes, radio luminosities (computed using integral fluxes for resolved sources and peak fluxes for unresolved sources, as described in Section \ref{sec-id}), and projected distance to each of the five clusters. There are 40 galaxies which meet the requirements outlined above, plus an additional seven for which velocities are not available. Of these seven, two are the aforementioned members of galaxy pairs whose partners have measured velocities placing them within the supercluster. As such, they are probable members of the supercluster. The remaining five galaxies include four potential cluster members and a probable background galaxy, evaluated on the basis of their optical colors. Analysis of radio galaxy fractions has been performed both with and without the six possible supercluster galaxies which lack direct velocity measurements (the presumed background galaxy was removed from consideration). As they are distributed across the supercluster, at most two such galaxies enter the calculation for any individual cluster. Consequently, they have little effect on the calculated fractions (see Tables \ref{tbl-brightfracs} - \ref{tbl-faintfracs}). The latter portion of Table \ref{tbl-abs} includes the same information for cluster radio galaxies which did not enter the radio galaxy fraction analysis.

Results of the radio galaxy fraction analysis are presented in Tables \ref{tbl-brightfracs}, \ref{tbl-intfracs}, and \ref{tbl-faintfracs} for the bright, intermediate, and faint optical magnitude bins. There is frequent evidence for significant variation in the radio galaxy fraction in the Shapley clusters. In general, there appears to be a deficit of radio galaxies in the intermediate optical magnitude bin (Table \ref{tbl-intfracs}). This holds for all five clusters at greater than the 95\% confidence level, although inclusion of the few galaxies without velocities reduces the significance below 95\% for A3562 and SC1329-313. The comparison sample indicates that about 15\% of galaxies in this optical magnitude range should be radio sources, while the values computed for individual clusters in the Shapley supercluster were all under 10\%, even when the radio galaxies lacking velocities were included. To assess the importance of cosmic variance in the computed significance levels, the fractions were recalculated under the assumption that all counted galaxies were cluster members (i.e., no background correction and hence a lower limit to the radio galaxy fraction) and using twice the assumed background correction. The resulting ranges for the calculated significance of any excess or deficit of radio galaxies are included in Tables \ref{tbl-brightfracs} - \ref{tbl-faintfracs}. It can be seen that the background correction does not appear to cause the deficit of intermediate optical magnitude radio galaxies.

The reduction in activity among intermediate optical magnitude galaxies is offset somewhat by an increase in the probability for the brightest galaxies to host radio sources, although number statistics limit the significance of this statement. The clusters generally have about 10 to 20 galaxies in this magnitude range, of which the comparison sample indicates that 24\% should be radio sources. In A3556, more than half of all bright galaxies are radio detections, translating to an increase in activity significant at about 99\% confidence. With the exception of SC1329-313, the other clusters have marginal excesses significant at about the 85\% level. These results are much more dependent on the background correction.

As with A2255, the most significant variation is found in the optically-fainter galaxies. The optical luminosity function insures that there are many such galaxies per cluster, and in general the fraction of these galaxies which are also radio emitters is quite small -- less than 2\%. Both A3556 and A3558 have radio galaxy fractions consistent with this figure, but A3562 has a very strong excess of radio galaxies among its fainter galaxy population with nearly 11\% of these galaxies being radio sources. The significance of this excess is well over 99.9\%, regardless of the background correction or whether a single galaxy without measured velocity is included. As would be expected based on the large overlap in areas covered by the 2 Mpc radial sampling limit (e.g., refer to Figure \ref{fig-foot}), SC1327-312 and SC1329-313 also have larger radio galaxy fractions than would be expected based on the comparison sample, at 98.5\% and 99.9\% significance respectively. The significance level for each of these three clusters would be higher if their allowed velocity ranges for radio galaxies were greater (i.e., as less rich systems within the supercluster they have lower $\sigma_v$ and consequently some radio galaxies with velocities consistent with supercluster membership are excluded from consideration as being outside the allowed range of $\pm3 \sigma_v$ for the individual clusters).

\subsubsection{Local Radio Galaxy Fractions Across the Supercluster}
An advantage to performing one study for the entire core of the supercluster is that any variations in radio galaxy fraction across the clusters may be assessed in light of environmental factors. To this end, a map showing the distribution of the radio galaxy fractions was created. At each point in the map the radio galaxy fraction was computed within a 0.5 Mpc radius, using a radio detection threshold of $6.8 \times 10^{21}$ W Hz$^{-1}$ and corrected galaxy counts with magnitudes $M_R \leq -20$ (i.e., luminosity and magnitude limits identical to the preceding analysis). The choice of a 0.5 Mpc sampling region was motivated by the desire to include large enough areas to have meaningful numbers of sources while still avoiding edge effects within the main body of the supercluster. Including regions beyond 2 Mpc of the respective cluster centers adds three additional radio galaxies assumed to be cluster members but for which no velocity information was available, and two further cases where radio detections without velocity measurements were assumed to be background sources on the basis of their red colors. Thus, contamination by foreground/background radio galaxies is still minimal. The resulting map is presented in Figure \ref{fig-fracN}, where the grey-scale indicates the radio galaxy fraction. To help indicate where high fractions result from low number statistics (e.g., only two local galaxies of which one is a radio source), contours of galaxy surface density are provided. The increase in activity in A3562 is readily apparent, particularly in the direction of A3558. 

The limits of $6.8 \times 10^{21}$ W Hz$^{-1}$ and $M_R \leq -20$ for the fractions analysis were set by the available comparison sample data. The present study includes much deeper data, so it is illustrative to consider fainter sources. Figure \ref{fig-fracT} parallels Figure \ref{fig-fracN} in that it is for sources with $M_R \leq -20$, but in this case radio sources with $2.1 \times 10^{21} \leq L_{1.4} \leq 6.8 \times 10^{21}$ W Hz$^{-1}$ are considered detections. The lower limit on radio luminosity corresponds to 0.5 mJy (i.e., 5$\sigma$ detections in the noisier regions of the mosaic) while the upper limit prevents direct overlap with the radio galaxies used in Figure \ref{fig-fracN}. Figure \ref{fig-fracP} is for the optically faintest galaxies in the present study, those with $-20 <  M_R \leq -19$. The radio detection limit is again $2.1 \times 10^{21}$ W Hz$^{-1}$; none of these fainter galaxies have $L_{1.4} > 6.8 \times 10^{21}$ W Hz$^{-1}$. As would be expected, most of the optically-fainter galaxies lack spectroscopic velocity measurements and are evaluated as cluster or non-cluster sources on the basis of $B - R$ color.

\section{Discussion}\label{sec-discuss}

\subsection{Implications of Radio Galaxy Fractions for Evolution}\label{ssec-evolution}

The identification of 210 radio galaxies in the core of the Shapley supercluster provides a good database for evaluation of galaxy activity. Only 34 of these radio galaxies may be removed from consideration as galaxies merely seen in projection on the supercluster, half of which are known foreground/background objects on the basis of existing optical spectroscopy with the other half appearing too red to be consistent with supercluster membership. Of 104 confirmed cluster members, at least 40 may be used for a statistical analysis of radio galaxy fractions which parallels that of MO03.

The analysis of radio galaxy fractions indicates varied activity across the supercluster. In general, there are two broad findings which need to be understood: 1) the decrease in the radio galaxy fraction for galaxies with optical magnitudes near $m^*$, generally seen across the entire core of the supercluster (i.e., the whole area studied); and 2) the strong increase in the radio galaxy fraction for optically-faint galaxies in the vicinity of A3562. These two points will now be discussed in light of environmental effects and prior theoretical and observational findings. 

Examination of the locations of the radio galaxies (Figures \ref{fig-fracN} - \ref{fig-fracP}) points to the region around A3562 as one of high activity. This is especially the case for the abundant optically-fainter galaxies, where the increase in the radio galaxy fraction for A3562 was significant at very high confidence. The fact that this increase corresponds to a specific region within the core of the supercluster indicates that there must be some difference in environment intrinsic to this region. In MO03, it was argued that A2255 had a larger active galaxy population as a result of an ongoing cluster-cluster merger. The A2255 merger axis is believed to be perpendicular to the observer and viewed very near the time when the cores of the merging partners are coincident. According to the simulations of \citet{roettiger1996}, at this particular time galaxies will cross the shock front formed between the merging clusters and experience a spike in ram pressure, thereby instigating starbursts. Recent X-ray analysis supports a similar picture in regards to the merger history and timing for A3562.

\citet{finoguenov2004} used a 6-pointing {\it XMM} mosaic to study the hydrodynamics of A3562 and the SC1329-313 group. They found an interesting assortment of properties, including temperature and entropy substructure and evidence for disruption of the core of SC1329-313. On $\sim$100 kpc scales, the emission of A3562 and SC1329-313 point towards one another, indicating interaction. The properties are well described by a scenario in which SC1329-313 has recently passed to the north of the core of A3562, traveling toward the west, and been deflected to the south and towards the observer. The motion was supersonic, having an implied Mach number $\approx$1.3. This passage would then have created a shock wave, as well as having produced oscillations of the core of A3562.

Examination of Figure \ref{fig-fracN} is qualitatively consistent with this scenario. The region associated with an excess of radio galaxies corresponds to the expected location of the shock associated with the passage of SC1329-313. In the \citet{roettiger1996} model, early in the merger the shock front acts as a buffer for any gas-rich galaxies in SC1329-313, protecting them from ram pressure removal of their gas. However, at a time around core passage the shock front decelerates and the galaxies pass through it, thereby experiencing a sharp increase in ram pressure leading to starbursts. An important consideration to this model for the Shapley supercluster, however, is the pre-merger nature of the active galaxies. Should they have been members of either A3562 or SC1329-313, the high efficiency of ram pressure stripping \citep[e.g.,][]{quilis2000} implies that it would be reasonable to expect that they were already gas poor and not likely to undergo a subsequent starburst. Consequently, the cluster merger itself may be indicative of larger-scale interactions: as the pair of clusters merge along a filament, outlying groups and field galaxies along the filament are also thrown into the mix. It is these galaxies which would cause the increase in the observed radio galaxy fraction. In a similar light, the merger may induce activity in the outlying groups through tidal interactions \citep{bekki1999,gnedin2003}. A picture along these same lines is evoked for the unusually high fraction of radio galaxies in A2125 \citep{owen2005b}, and seems to fit for the curious radio galaxy J133048-314325 in the Shapley supercluster.

Figure \ref{fig-133048} depicts J133048-314325, which superficially has the radio morphology of a wide-angle tail (WAT) but a radio luminosity a factor of ten lower than such sources. Two other galaxies are easily visible nearby this source, the radio galaxy J133051-314417 and a face-on spiral galaxy (J133047-314339) within the radio contours of J133048-314325 \citep[note that this spiral galaxy was designated the radio counterpart in lower resolution observations by ][]{giacintucci2004}. These three galaxies appear to be a compact group, as indicated by their small spatial and velocity separation (a maximum separation of just 1\arcmin, or about 50 kpc; NED velocities of 13394, 13226, and 13815 \kms{} in order of increasing galaxy RA). A potentially useful analogy for this system is that of Stephan's Quintet in which a high velocity intruder galaxy (NGC~7318A) has interacted with the intragroup medium (IGM) around NGC~7318A and NGC~7319. Radio continuum emission from this system \citep[e.g.,][]{xu2003} arises from a combination of sources, including the Seyfert nucleus of NGC~7319, star formation, and the large ($\sim40$ kpc) shock caused by the interaction. Much of the star formation, interestingly enough, is within the IGM and not simply associated with individual galaxies of Stephan's Quintet. While this star formation does contribute to the 1.4~GHz radio continuum, the majority of the radio emission is associated with a radio ridge delineating the shock front (about 1.5 mJy for star formation, 35 mJy for the shock, and 29 mJy for the Seyfert galaxy NGC~7319). Inspection of the radio contours of J133048-314325 suggests a similar interaction of the sources, potentially explaining their activity. J133051-314417 would be the intruder galaxy, with support for this interpretation being its $\sim500$ \kms{} velocity offset from the other galaxies and the general shape of the radio emission which is suggestive of J133051-314417 having passed through the IGM in a southeasterly direction. The extended morphology of J133048-314325 would then be the combination of possible AGN emission from the galaxy itself, plus shock and star formation emission. That J133048-314325 might be an AGN is suggested by its optical properties, which include what might be an unresolved optical nucleus coincident with the peak in the radio emission, its blue color $B - R = 1.11$, and an unusually bright absolute magnitude of $M_R = -23.7$. It is possible that the unresolved optical component and bright total magnitude arise from a foreground star, as no emission lines are noted in the spectroscopy reported by \citet{stein1996}, the literature source for the NED velocity. However, it would be a remarkable coincidence if a foreground star were to lie in projection directly on the nucleus of a galaxy which has moderately strong radio emission. The portion of the J133048-314325 emission that would be associated with the shock would then be the ridge to the southeast. Using TVSTAT to evaluate the flux associated with this feature produces a radio luminosity of $\sim1.5 \times 10^{22}$ W Hz$^{-1}$, comparable to that of the shock ridge in Stephan's Quintet ($L_{1.4} = 2.9 \times 10^{22}$ W Hz$^{-1}$). Its size, $\sim$60 kpc, is also similar to that of the shock ridge in Stephan's Quintet. A contribution to the total emission along the northwest edge could then come from star formation in the spiral galaxy J133047-314339. The entire system lies about 11\arcmin{} to the northwest of the cataloged position of SC1329-313, very near the peak in the galaxy surface density indicated in Figure \ref{fig-fracN}. Thus, it is reasonable to think that the hypothesized interaction akin to that seen in Stephan's Quintet is the result of tidal forces from the larger merger of SC1329-313 with A3562.

This hypothesis can easily be tested by future observations. Additional radio observations at shorter wavelengths would reveal any spectral index variations, where any shock emission would have steeper spectral index \citep[e.g., the shock ridge in Stephan's Quintet has $\alpha=0.93\pm0.13$][]{xu2003}. Higher resolution observations might also help separate any AGN emission associated with J133048-314325. Finally, \citet{xu2003} use infrared observations and long-slit optical spectroscopy to reveal regions associated with shocked gas and those associated with photoionization from star formation.

Returning to our discussion of radio galaxy fractions, there is less direct evidence for the ``preferred'' location of activity around SC1329-313 when examining fractions at fainter optical and radio limits. However, these results may still be interpretted within the framework described above. The inclusion of galaxies with fainter radio luminosities while keeping the optical limit at $M_R \leq -20$ (Figure \ref{fig-fracT}) produces a more uniform distribution of radio galaxy activity within the supercluster. For the faintest optical galaxies studied, those with $M_R$ between -19 and -20, the radio sources are again found mainly within A3562 (Figure \ref{fig-fracP}). In each of these cases, the star formation rates implied by the radio emission are $\approx1 - 4$ M$_\odot$ yr$^{-1}$ \citep{yun2001}. This would be a substantial amount of star formation in the fainter galaxies (i.e., those with $M_R \lesssim -20$), and a typical star formation rate for normal brighter spirals such as the Milky Way. Thus, it again appears that the optically fainter galaxies are likely to be starbursts, and that these types of galaxies are found preferentially in the vicinity of A3562. 

Why would there be different effects on the faint and intermediate magnitude populations? The answer to this question might lie in the different galaxy morphologies that are included within these magnitude ranges. Galaxies in the intermediate magnitude range will be larger and more likely to have significant bulges. It is a common prediction of evolutionary models that bulges stabilize galaxies against transformation \citep[e.g., galaxy-galaxy mergers and harassment,][]{mihos1996,moore1998}. This is because the predicted evolution is usually driven by the rapid inflow of gas to the nuclei of galaxies, where it creates a nuclear starburst. Thus, brighter galaxies with bulges are more resistant to evolutionary changes than fainter, disk-dominated galaxies. Although this can easily explain why more dramatic change is seen in the fainter galaxy population, it is still difficult to understand why there would be a deficit of intermediate magnitude radio galaxies in the Shapley supercluster relative to other clusters. It is possible that the radio galaxies in this magnitude range found in other clusters represent the steady trickle of isolated field galaxies that enter the clusters, and that the dramatic merger environment of the Shapley supercluster precludes such objects.

\subsection{Comparison to Prior Radio Studies}

Are these results consistent with the prior radio studies of the core of the Shapley supercluster? \citet[][hereafter V00]{venturi2000} examined A3558 and argued for a reduction in galaxy activity in this region, and \citet[][G04]{giacintucci2004} studied A3562 and found a level of activity normal for clusters in general.

V00 noted a reduced fraction of radio galaxies in A3558 relative to the cluster survey of \citet[][LO96]{ledlow1996}. In the present study, there is a reduced fraction of radio galaxies in the intermediate optical magnitude range for A3558 while the fractions among the brighter and fainter optical magnitudes are normal. In Figure \ref{fig-rlf}, the cumulative radio luminosity function (RLF) for the Shapley supercluster and that for A3558 alone are plotted against the RLF derived for the comparison sample from \citet{miller2001}. In each case, the RLF is for galaxies with $M_R \leq -20$ and within 2 Mpc projected separation of their cluster centers (hence the A3558 RLF is a subset of the presented Shapley supercluster RLF). The LO96 RLF was constructed using only elliptical galaxies with $M_R \leq -20.5$, so it is not immediately comparable to those in the figure. However, applying a simple scaling which amounts to saying that 13.5\% of all cluster galaxies with $M_R \leq -20$ are ellipticals with $M_R \leq -20.5$ produces an excellent agreement with the bright end of the RLF derived from the \citet{miller2001} comparison sample. Since essentially all radio galaxies with $L_{1.4} \geq 10^{23}$ W Hz$^{-1}$ are bright ellipticals, this agreement is expected.

As can be seen in Figure \ref{fig-rlf}, there does appear to be a deficit of the more luminous radio galaxies in A3558 and in the Shapley supercluster core region in general. This might have been expected based on the result of a lower fraction of intermediate optical magnitude radio galaxies, although that finding was irrespective of radio luminosity and was countered slightly by the increased fraction of optically bright radio galaxies. The difficulty in assessing any deficit of strong radio galaxies relative to the LO96 RLF is that galaxies with such high luminosities are intrinsically rare, so demonstrating a statistically significant absence for a specific cluster is hampered by number statistics. In their analysis, V00 attempted to perform a more direct comparison with the LO96 RLF by correcting total galaxy counts to estimate the number of ellipticals and S0s in the A3558 environment. They arrived at a net of 185 elliptical and S0 galaxies, of which 17 were identified radio galaxies. Integration of the LO96 RLF would predict 26, and thus the deficit appears significant at about 97\% confidence. However, the LO96 RLF was constructed for ellipticals only and LO96 noted that only  $\sim$3 of their sample of 188 radio sources were S0 galaxies with the remainder being ellipticals. Should just 24 of the 185 estimated counts used in the V00 analysis be S0 galaxies, the significance of the deficit drops under 90\%.

It is instructive to examine this issue more closely. Following the LO96 prescription, all galaxies within 0.6 Mpc of the center of A3558 were selected (i.e., 0.3$R_{Abell}$). These were further culled to remove galaxies with $M_R > -20.5$ and $B - R < 1.46$, for more direct comparison with the V00 analysis. This produced a list of 43 galaxies, of which three had radio luminosities high enough for consideration. A simple integration of the LO96 RLF would predict about 6 radio galaxies, and thus yields a deficit that would be claimed with 91\% significance. The morphologies of the 43 selected galaxies were assessed visually using the $R$ images along with ellipticities determined by SExtractor. For most of these galaxies, NED also provided a morphology (including confirmation of cluster membership via a published velocity) and the assignments were consistent. Removal of non-ellipticals left only 24 galaxies, including the three radio galaxies. This is almost exactly the fraction predicted by the LO96 RLF. This analysis underscores the uncertainties involved in comparing luminosity functions constructed using small numbers of galaxies. Note that although the preceding discussion was based on photometry, the alternate procedure of V00 using spectroscopy is similarly effected: selecting galaxies on the basis of non-emission line spectra will include S0 galaxies as well as ``passive spirals'' \citep{bekki2002,goto2003} and potentially even normal spirals whose fiber spectra sample just their bulge components. In summary, although there may be a reduction in the fraction of A3558's elliptical galaxy population which host radio sources, it is questionable whether this reduction is significant.

G04 noted only marginal evidence for an increase in the radio-emitting faint galaxy population in A3562, whereas in the current work it is reported as highly significant. Similarly, their analysis of the data presented in MO03 suggested that the radio galaxy fraction in A2255 was only marginally greater than the comparison clusters. This difference in interpretation is caused primarily by the adopted statistical methods. Using the same magnitude ranges described above, G04 calculated the fraction of radio galaxies for each cluster. They then calculated the fraction for the full comparison sample as the mean of these values, with the standard deviation representing the error.\footnote{A more appropriate error for the comparison case would be the standard deviation of the mean, i.e. the dispersion divided by the square root of the number of clusters. The error for any individual cluster would be represented by the dispersion of the full sample.} The procedure of MO03 (and used here) treats the comparison sample as a whole. In essence, it is an application of the binomial distribution where the comparison sample is used to determine the expected probability. Using numbers for faint galaxies from Table \ref{tbl-faintfracs}, 20 of 1104.6 comparison sample galaxies were radio detections so the statistical question is: what is the probability that finding 11 of 113.2 galaxies (i.e., the results for A3562) is consistent with this parent distribution?  The $\chi ^2$ nature of the actual test takes into account that the ``true fraction,'' i.e., the 20 of 1104.6 from the comparison sample, is not known with certainty as would be assumed if one used the standard binomial probability function. These differences in applied statistical tests are the reason that G04 ascribed only marginal significance to the excess activity in A2255. Note that in the case of A3562, the G04 numbers for radio detections and total galaxies do not produce a significant result regardless of the statistical method used. Presumably this is the result of differences in the areas surveyed. 

\subsection{Initiated Starbursts or Strangled Field Galaxies?}

Although the radio data presented herein identify the region around SC1329-313 to have an excess of active galaxies, they only indirectly address the level of this activity. The important question with respect to cluster evolutionary models, particularly those applied to the Butcher-Oemler effect, is whether the active galaxies are starbursts or more normal star-forming galaxies which have just entered the supercluster. In some studies, clusters of galaxies at moderate redshifts ($z \sim 0.4$) show evidence for starbursts \citep[e.g.,][]{poggianti1999} while other studies suggest star formation is gradually extinguished as stripping removes the gaseous halos from galaxies, thereby removing the reservoir needed for future star formation \citep[``strangulation,'' e.g.,][]{balogh1999}. 

Unfortunately, the data presented herein do not allow for straightforward testing of such issues as the radio emission may include an AGN component. In the case of the optically fainter galaxies ($M_R > -20$), a good case can be made for starbursts on the basis of their relatively high radio luminosities and the fact that AGN are rare in low mass galaxies. For the brighter radio galaxies, ideally one would use optical spectroscopy in comparison with models \citep[as done in ][]{poggianti1999} or data at other wavelengths such as the UV in order to quantify the past-to-present star formation and fraction of total stellar mass created in recent starbursts \citep[e.g.,][]{kauffmann2003,salim2005}. As alluded to in the case of the fainter galaxies, one possible available avenue is to rely on the strength of the radio emission to indicate the level of present star formation and the $R$ magnitude as a proxy for total stellar mass. The ratio of these quantities would then be a measure of the relative strength of the current star formation. 

The radio-to-optical flux ratio, $r$, was calculated using the definition presented in \citet{machalski1999}: $r = \log (S_{1.4}/\mathnormal{f}_R)$, where $\mathnormal{f}_R$ is the flux density at 6940$\mbox{\AA}$ determined using the $R$ magnitude via $\mathnormal{f}_R = 2.78 \times 10^{6 - 0.4R}$. To reduce contamination due to AGN, galaxies brighter than $M_R = -22$ were removed along with those redder than $B - R = 1.45$ (i.e., remove probable red sequence objects down to the fainter magnitudes studied). In addition a minimum 1.4~GHz flux of 0.5 mJy was required to remove potential bias caused by variation in the sensitivity of the radio data across the supercluster. After removal of galaxies with published velocities placing them outside the supercluster, this resulted in a set of 83 galaxies for testing (48 of which did have published velocities indicating supercluster membership). Specific regions were compared via a Wilcox test to determine whether any exhibited evidence for higher $r$ values, and hence more likely greater relative levels of star formation. No highly significant variation in $r$ values across the core of the supercluster was found, although there is slight evidence that those in the vicinity of SC1329-313 do have higher $r$. This result was always less than $2\sigma$, and slightly dependent on sub-samples used (e.g., how the regions were defined in RA and Dec, whether or not galaxies without published velocities were included, what optical magnitude ranges were compared, etc.). Hence, this simple statistical analysis is suggestive of increased starburst activity associated with SC1329-313, although this clearly needs to be confirmed through a more direct study.

\section{Conclusions}

This paper has presented a comprehensive radio study of the core of the Shapley supercluster. The VLA has been used to map a nearly 7 square degree area through a mosaic of pointings which provide moderately uniform noise characteristics across the entire area. In conjunction with optical imaging, radio detections for 210 galaxies with $m_R \leq 17.36$ were presented. These include 104 galaxies with velocities placing them within the supercluster, which approximately doubles the previously known total of radio galaxies in the core of the Shapley supercluster. In addition, 2 radio galaxies are members of cataloged pairs whose companions have supercluster velocities and 8 radio galaxies are likely associated with the supercluster but are formally placed in the foreground (9000 \kms{} $< cz <$ 10934 \kms). Of those radio galaxies without velocity measurements, 68 have optical colors which do not rule out supercluster membership.

Across the entire core of the supercluster, intermediate optical magnitude ($-22 < M_R \leq -21$) galaxies appear less likely to host radio sources than their counterparts in a large comparison sample. About 5\% of such galaxies were found to be radio sources with $L_{1.4} \geq 6.8 \times 10^{21}$ W Hz$^{-1}$, whereas 15\% of the comparison sample were. This deficit is offset somewhat by the brighter galaxies ($M_R \leq -22$) of the Shapley supercluster being more likely to be radio sources.

While these results generally pertained to the entire region surveyed, a more dramatic effect was observed in the fainter galaxies ($-21 < M_R \leq -20$) localized around A3562 and SC1329-313. These galaxies were much more likely to be radio sources than anticipated on the basis of comparison clusters, with the high significance of the result being unaffected by changes in the assumed background. On the basis of their blue colors and radio luminosities, these galaxies are presumably starbursts. It is fascinating that this very region is identified by recent X-ray analysis as the location of an ongoing merger of SC1329-313 with A3562 and the rest of the supercluster. The remainder of the supercluster does not exhibit any statistical excess of radio galaxies among its optically fainter population, underscoring the potential importance of cluster mergers in galaxy evolution. This interpretation is consistent with radio studies of other active clusters, notably A2125 and A2255.

Examination of the galaxies with extended radio emission also revealed two interesting candidates for future observation. J132802-314521 is a strong radio source with a luminous core coincident with the galaxy ($L_{1.4} > 10^{23}$ W Hz$^{-1}$). This emission appears to connect to larger-scale diffuse emission, suggesting this source might be another head-tail radio galaxy within the supercluster. Extended radio emission around the galaxy J133048-314325 is a potential analog to the well-known system Stephan's Quintet. In this interpretation, J133051-314417 would have interacted with J133048-314325 and the nearby spiral J133047-314339. The resulting radio emission would be the combination of emission from the individual galaxies as well as shocks in the IGM. Interestingly, this source is near the peak of the galaxy distribution centered on SC1329-313, suggesting that environmental processes in groups involved with larger-scale mergers may be important for galaxy evolution.

\acknowledgments
The author thanks an anonymous referee, whose careful reading of the manuscript led to improvements in the clarity of the statistical analysis and additional insight into the extended radio galaxies. Much of this work was completed while I held a National Research Council Associateship at NASA's Goddard Space Flight Center. I also acknowledge the support of NASA through the American Astronomical Society's Small Research Grant Program, which supplied the computing resources to analyze the radio data. Most significantly, this project benefitted greatly from the assistance of Michael Ledlow. Mike's enthusiastic support of the project included invaluable comments on observing proposals and assistance with the data collection. I am unable to express enough how sad his passing was, and how great a loss it is to his friends and colleagues.

\clearpage

\begin{deluxetable}{r r r c c}
\tablecolumns{5}
\tablecaption{Summary of Radio Observations\label{tbl-radobs}}
\tablewidth{0pt}
\tablehead{
\colhead{Field ID} & \colhead{RA(J2000)} & \colhead{Dec(J2000)} & 
\colhead{Beam Size\tablenotemark{a}} & \colhead{Dates\tablenotemark{b}}
}
\startdata
 1A & 13 28 48.0 & -30 41 00 & 23 $\times$ 11 & 06/18/2001 \\
 2B & 13 27 35.4 & -30 50 00 & 20 $\times$ 12 & 06/18/2001 \\
 2C & 13 30 00.6 & -30 50 00 & 26 $\times$ 10 & 06/20/2001 \\
 3A & 13 23 57.1 & -30 58 45 & 37 $\times$ \phn7 & 06/21/2001 \\
 3B & 13 26 22.5 & -30 58 58 & 29 $\times$ \phn9 & 06/22/2001 \\
 3C & 13 28 48.0 & -30 59 00 & 19 $\times$ 14 & 06/18/2001 \\
 3D & 13 31 13.5 & -30 58 58 & 20 $\times$ 12 & 06/20/2001 \\
 4A & 13 22 43.8 & -31 07 34 & 28 $\times$ \phn9 & 06/18/2001 \\
 4B & 13 25 09.5 & -31 07 52 & 29 $\times$ \phn9 & 06/21/2001 \\
 4C & 13 27 35.2 & -31 08 00 & 17 $\times$ 14 & 06/18/2001 \\
 4D & 13 30 00.8 & -31 08 00 & 18 $\times$ 14 & 06/18/2001 \\
 4E & 13 32 26.5 & -31 07 52 & 20 $\times$ 13 & 06/20/2001 \\
    &            &           &                & 06/22/2001 \\
 4F & 13 34 52.2 & -31 07 34 & 19 $\times$ 13 & 06/22/2001 \\
 5A & 13 21 30.3 & -31 16 21 & 27 $\times$ \phn9 & 06/17/2001 \\
 5B & 13 23 56.1 & -31 16 45 & 24 $\times$ 10 & 06/17/2001 \\
 5C & 13 26 22.1 & -31 16 58 & 21 $\times$ 12 & 09/29/2002 \\
 5D & 13 28 48.0 & -31 17 00 & 17 $\times$ 15 & 06/18/2001 \\
 5E & 13 31 13.9 & -31 16 58 & 16 $\times$ 15 & 06/20/2001 \\
    &            &           &                & 06/22/2001 \\
 5F & 13 33 39.9 & -31 16 45 & 18 $\times$ 15 & 06/22/2001 \\
 5G & 13 36 05.7 & -31 16 21 & 16 $\times$ 16 & 06/22/2001 \\
 6A & 13 22 42.6 & -31 25 34 & 23 $\times$ 11 & 06/17/2001 \\
 6B & 13 25 08.8 & -31 25 52 & 18 $\times$ 14 & 06/17/2001 \\
 6C & 13 27 34.9 & -31 26 00 & 17 $\times$ 16 & 06/17/2001 \\
    &            &           &                & 09/29/2002 \\
 6D & 13 30 01.1 & -31 26 00 & 19 $\times$ 13 & 06/18/2001 \\
 6E & 13 32 27.2 & -31 25 52 & 16 $\times$ 15 & 06/22/2001 \\
    &            &           &                & 09/29/2002 \\
 6F & 13 34 53.4 & -31 25 34 & 17 $\times$ 15 & 06/22/2001 \\
 7A & 13 21 28.8 & -31 34 21 & 20 $\times$ 13 & 06/17/2001 \\
 7B & 13 23 55.2 & -31 34 45 & 17 $\times$ 14 & 06/17/2001 \\
 7C & 13 26 21.6 & -31 34 58 & 16 $\times$ 16 & 09/29/2002 \\
 7D & 13 28 48.0 & -31 35 00 & 16 $\times$ 16 & 06/18/2001 \\
 7E & 13 31 14.4 & -31 34 58 & 18 $\times$ 14 & 06/22/2001 \\
    &            &           &                & 09/29/2002 \\
 7F & 13 33 40.8 & -31 34 45 & 16 $\times$ 15 & 06/20/2001 \\
    &            &           &                & 09/29/2002 \\
 7G & 13 36 07.2 & -31 34 21 & 17 $\times$ 15 & 06/22/2001 \\
 8A & 13 20 14.8 & -31 43 05 & 18 $\times$ 14 & 06/17/2001 \\
 8B & 13 22 41.4 & -31 43 34 & 17 $\times$ 15 & 06/17/2001 \\
 8C & 13 25 08.1 & -31 43 52 & 17 $\times$ 15 & 09/29/2002 \\
 8D & 13 27 34.7 & -31 44 00 & 18 $\times$ 14 & 09/29/2002 \\
 8E & 13 30 01.3 & -31 44 00 & 17 $\times$ 15 & 09/29/2002 \\
 8F & 13 32 28.0 & -31 43 52 & 17 $\times$ 15 & 06/20/2001 \\
    &            &           &                & 09/29/2002 \\
 8G & 13 34 54.6 & -31 43 34 & 18 $\times$ 14 & 06/22/2001 \\
 9A & 13 21 27.4 & -31 52 21 & 17 $\times$ 15 & 06/17/2001 \\
 9B & 13 23 54.2 & -31 52 45 & 18 $\times$ 14 & 09/29/2002 \\
 9C & 13 26 21.1 & -31 52 58 & 19 $\times$ 13 & 09/29/2002 \\
 9D & 13 28 48.0 & -31 53 00 & 21 $\times$ 12 & 06/18/2001 \\
 9E & 13 31 14.9 & -31 52 58 & 25 $\times$ 10 & 06/22/2001 \\
 9F & 13 33 41.8 & -31 52 45 & 20 $\times$ 13 & 09/29/2002 \\
 9G & 13 36 08.6 & -31 52 21 & 20 $\times$ 13 & 06/22/2001 \\
10A & 13 22 40.2 & -32 01 34 & 16 $\times$ 16 & 06/17/2001 \\
10B & 13 25 07.3 & -32 01 52 & 25 $\times$ 10 & 09/29/2002 \\
10C & 13 27 34.4 & -32 02 00 & 17 $\times$ 15 & 06/18/2001 \\
10D & 13 30 01.6 & -32 02 00 & 24 $\times$ 11 & 06/18/2001 \\
10E & 13 32 28.7 & -32 01 52 & 25 $\times$ 10 & 06/22/2001 \\
10F & 13 34 55.8 & -32 01 34 & 22 $\times$ 12 & 06/22/2001 \\
11A & 13 21 25.9 & -32 10 21 & 20 $\times$ 13 & 06/17/2001 \\
11B & 13 23 53.3 & -32 10 45 & 29 $\times$ \phn9 & 06/17/2001 \\
11C & 13 26 20.6 & -32 10 58 & 30 $\times$ \phn9 & 09/29/2002 \\
11D & 13 28 48.0 & -32 11 00 & 28 $\times$ \phn9 & 06/18/2001 \\
11E & 13 31 15.4 & -32 10 58 & 35 $\times$ \phn7 & 09/29/2002 \\
11F & 13 33 42.7 & -32 10 45 & 33 $\times$ \phn8 & 06/20/2001 \\
12A & 13 22 39.0 & -32 19 34 & 22 $\times$ 12 & 06/17/2001 \\
12D & 13 30 01.8 & -32 20 00 & 34 $\times$ \phn8 & 06/18/2001 \\
12E & 13 32 29.4 & -32 19 52 & 21 $\times$ 12 & 06/20/2001 \\
12F & 13 34 57.0 & -32 19 34 & 26 $\times$ 10 & 06/21/2001 \\
\enddata

\tablenotetext{a}{In the final imaging, all beams were restored with a 16\arcsec{} $\times$ 16\arcsec{} Gaussian.}

\tablenotetext{b}{All of the 2001 June observations were performed at night, minimizing solar interference, while the 2002 September observations occurred during daytime hours. Some data were lost due to strong winds stowing the array on 06/20/2001, and only a few pointings were successfully completed on 06/21/2001 due to mechanical problems.}

\end{deluxetable}

\clearpage

\begin{deluxetable}{r r r r r r r c r r r c}
\tablecolumns{12}
\tablecaption{Radio-Detected Galaxies with $m_R \leq 17.36$\label{tbl-radgals}}
\tabletypesize{\scriptsize}
\tablewidth{0pt}
\tablehead{
\colhead{RA} & \colhead{Dec} & \colhead{$m_R$} & \colhead{$m_B$} & \colhead{$B - R$} & 
\colhead{$S_I$} & \colhead{$S_P$} & \colhead{Res?} & \colhead{rms} & \colhead{Sep} &
\colhead{Velocity} & \colhead{Comments}
}
\startdata
13 19 03.14 & -31 34 42.9 & 16.42$\pm$0.10 & 19.33$\pm$0.25 & 2.91$\pm$0.27 &  1.50 & 0.61 & R & 0.20 & 4.8 & \nodata & a \\
13 19 07.85 & -31 40 10.8 & 15.39$\pm$0.10 & 16.91$\pm$0.05 & 1.52$\pm$0.11 &  0.94 & 0.67 & U & 0.14 & 5.2 & 13968 & \nodata \\
13 19 15.67 & -31 51 18.8 & 16.87$\pm$0.11 & 18.88$\pm$0.12 & 2.01$\pm$0.16 &  5.14 & 3.40 & R & 0.15 & 4.5 & \nodata & a \\
13 19 35.06 & -31 31 39.4 & 16.30$\pm$0.10 & 18.21$\pm$0.09 & 1.91$\pm$0.13 &  1.76 & 1.73 & R & 0.13 & 2.3 & \nodata & a \\
13 19 35.35 & -31 40 21.3 & 15.39$\pm$0.10 & 16.46$\pm$0.05 & 1.07$\pm$0.11 &  0.63 & 0.70 & U & 0.09 & 2.9 & \nodata & \nodata \\
13 20 23.40 & -31 26 31.8 & 16.41$\pm$0.10 & 18.40$\pm$0.16 & 1.99$\pm$0.19 &  1.01 & 0.45 & R & 0.11 & 4.3 & \nodata & a \\
13 20 24.47 & -32 02 48.0 & 16.89$\pm$0.11 & 19.86$\pm$0.61 & 2.97$\pm$0.62 &  1.00 & 0.70 & R & 0.12 & 1.1 & \nodata & a \\
13 20 27.89 & -31 36 07.1 & 16.85$\pm$0.11 & 18.45$\pm$0.14 & 1.60$\pm$0.18 &  0.48 & 0.36 & R & 0.09 & 2.7 & \nodata & \nodata \\
13 20 54.63 & -31 44 33.4 & 15.28$\pm$0.10 & 16.90$\pm$0.06 & 1.62$\pm$0.12 &  7.94 & 7.12 & R & 0.07 & 1.2 & 16249 & \nodata \\
13 20 55.59 & -31 43 59.3 & 14.94$\pm$0.10 & 15.91$\pm$0.05 & 0.97$\pm$0.11 &  6.65 & 5.12 & R & 0.07 & 2.1 & 15829 & \nodata \\
13 20 55.64 & -31 27 24.6 & 17.22$\pm$0.11 & 19.03$\pm$0.22 & 1.81$\pm$0.25 &  0.84 & 0.75 & R & 0.09 & 0.8 & \nodata & \nodata \\
13 21 12.30 & -31 08 14.9 & 15.45$\pm$0.10 & 17.17$\pm$0.06 & 1.72$\pm$0.12 &  1.33 & 1.58 & U & 0.12 & 1.1 & \nodata & \nodata \\
13 21 45.02 & -31 03 00.6 & 14.56$\pm$0.10 & 16.36$\pm$0.05 & 1.80$\pm$0.11 &  7.80 & 4.93 & R & 0.12 & 1.1 & 15071 & b,c \\
13 21 52.65 & -31 45 54.4 & 14.79$\pm$0.10 & 16.13$\pm$0.05 & 1.34$\pm$0.11 &  1.44 & 1.03 & R & 0.07 & 3.5 & 14705 & \nodata \\
13 21 58.97 & -31 38 37.8 & 15.40$\pm$0.10 & 16.90$\pm$0.06 & 1.50$\pm$0.12 &  0.45 & 0.45 & U & 0.07 & 2.5 & \nodata & \nodata \\
13 22 01.07 & -31 26 44.6 & 17.14$\pm$0.11 & 18.17$\pm$0.11 & 1.03$\pm$0.16 &  0.28 & 0.40 & U & 0.08 & 2.2 & \nodata & \nodata \\
13 22 03.25 & -31 02 24.7 & 17.35$\pm$0.12 & 19.14$\pm$0.22 & 1.79$\pm$0.25 &  0.77 & 0.58 & R & 0.11 & 4.7 & \nodata & \nodata \\
13 22 06.75 & -31 46 16.4 & 14.24$\pm$0.10 & 16.20$\pm$0.05 & 1.96$\pm$0.11 &  7.34 & 2.29 & R & 0.07 & \nodata & 14229 & b,c \\
13 22 10.13 & -31 40 52.1 & 16.97$\pm$0.10 & 17.81$\pm$0.08 & 0.84$\pm$0.13 &  1.85 & 0.96 & R & 0.07 & 3.0 & \nodata & \nodata \\
13 22 10.92 & -31 28 05.1 & 16.78$\pm$0.10 & 18.7\phn$\pm$0.2\phn & 1.9\phn$\pm$0.2\phn & 98.87 & 22.00 & R & 0.07 & \nodata & \nodata & a,b,c \\
13 22 19.65 & -31 22 05.7 & 16.00$\pm$0.10 & 17.06$\pm$0.06 & 1.06$\pm$0.12 &  0.55 & 0.72 & U & 0.07 & 3.0 & \nodata & \nodata \\
13 22 22.85 & -31 37 43.6 & 16.00$\pm$0.10 & 17.11$\pm$0.06 & 1.11$\pm$0.12 &  1.28 & 1.09 & R & 0.07 & 0.2 & \nodata & \nodata \\
13 22 23.73 & -31 04 32.7 & 13.97$\pm$0.10 & 15.22$\pm$0.05 & 1.25$\pm$0.11 &  2.38 & 0.91 & R & 0.09 & 1.7 & 4512 & \nodata \\
13 22 39.55 & -32 15 20.1 & 17.00$\pm$0.11 & 19.15$\pm$0.28 & 2.15$\pm$0.30 &  1.11 & 1.06 & U & 0.08 & 2.9 & \nodata & a \\
13 22 41.29 & -32 18 50.6 & 16.04$\pm$0.10 & 18.21$\pm$0.09 & 2.17$\pm$0.13 &  0.50 & 0.40 & R & 0.09 & 3.0 & \nodata & a \\
13 22 42.81 & -32 23 42.2 & 15.00$\pm$0.10 & 16.28$\pm$0.05 & 1.28$\pm$0.11 &  1.54 & 0.84 & R & 0.11 & 2.7 & \nodata & \nodata \\
13 22 49.71 & -32 26 11.6 & 14.31$\pm$0.10 & 16.05$\pm$0.05 & 1.74$\pm$0.11 &  2.92 & 3.18 & U & 0.12 & 3.7 & 13731 & \nodata \\
13 22 52.61 & -32 30 05.6 & 17.07$\pm$0.12 & 19.70$\pm$0.35 & 2.63$\pm$0.37 &  1.11 & 0.85 & R & 0.15 & 1.1 & \nodata & a \\
13 23 00.09 & -31 23 13.9 & 15.03$\pm$0.10 & 16.13$\pm$0.05 & 1.10$\pm$0.11 &  1.17 & 0.84 & R & 0.07 & 1.4 & \nodata & \nodata \\
13 23 03.86 & -31 23 59.1 & 16.50$\pm$0.11 & 17.95$\pm$0.06 & 1.45$\pm$0.13 &  2.61 & 1.27 & R & 0.06 & 2.4 & \nodata & \nodata \\
13 23 15.11 & -31 27 36.0 & 14.09$\pm$0.10 & 15.21$\pm$0.05 & 1.12$\pm$0.11 &  5.18 & 4.02 & R & 0.07 & 1.4 & 9068 & \nodata \\
13 23 29.07 & -31 50 39.6 & 13.86$\pm$0.10 & 15.58$\pm$0.05 & 1.72$\pm$0.11 &  4.25 & 3.85 & R & 0.08 & 1.7 & 14306 & \nodata \\
13 23 33.56 & -32 16 03.8 & 16.96$\pm$0.11 & 18.23$\pm$0.13 & 1.27$\pm$0.17 &  0.47 & 0.55 & R & 0.08 & 4.7 & \nodata & \nodata \\
13 23 54.22 & -32 10 56.3 & 14.32$\pm$0.10 & 15.64$\pm$0.05 & 1.32$\pm$0.11 &  5.76 & 3.05 & R & 0.08 & 1.3 & 15295 & \nodata \\
13 23 57.58 & -31 38 45.2 & 13.93$\pm$0.10 & 15.63$\pm$0.05 & 1.70$\pm$0.11 & 55.54 & 2.43 & R & 0.08 & 3.2 & 15054 & b,c \\
13 24 02.09 & -31 19 13.8 & 16.24$\pm$0.10 & 17.21$\pm$0.05 & 0.97$\pm$0.11 &  0.53 & 0.50 & R & 0.07 & 1.6 & \nodata & \nodata \\
13 24 05.04 & -31 04 19.6 & 16.92$\pm$0.11 & 18.68$\pm$0.07 & 1.76$\pm$0.13 &  1.11 & 1.26 & U & 0.08 & 4.2 & \nodata & a \\
13 24 06.68 & -31 40 12.0 & 13.49$\pm$0.10 & 15.28$\pm$0.05 & 1.79$\pm$0.11 & 10.39 & 9.06 & R & 0.08 & 0.9 & 14397 & \nodata \\
13 24 16.30 & -32 12 54.6 & 16.59$\pm$0.10 & 17.68$\pm$0.06 & 1.09$\pm$0.12 &  0.44 & 0.56 & R & 0.09 & 4.6 & \nodata & \nodata \\
13 24 16.88 & -31 10 09.9 & 14.47$\pm$0.10 & 16.03$\pm$0.05 & 1.56$\pm$0.11 &  2.04 & 1.66 & R & 0.08 & 2.5 & 10057 & \nodata \\
13 24 22.77 & -31 42 23.7 & 13.85$\pm$0.10 & 15.56$\pm$0.05 & 1.71$\pm$0.11 & 13.90 & 9.83 & R & 0.09 & 0.7 & 14370 & \nodata \\
13 24 32.46 & -31 06 58.2 & 15.65$\pm$0.10 & 16.81$\pm$0.05 & 1.16$\pm$0.11 &  2.50 & 2.07 & R & 0.08 & 1.0 & \nodata & \nodata \\
13 24 33.93 & -31 40 14.6 & 17.09$\pm$0.11 & 17.90$\pm$0.06 & 0.81$\pm$0.13 &  0.43 & 0.50 & U & 0.09 & 3.0 & \nodata & \nodata \\
13 24 37.96 & -31 52 39.0 & 17.05$\pm$0.11 & 18.88$\pm$0.11 & 1.83$\pm$0.16 &  0.73 & 0.55 & R & 0.10 & 1.7 & \nodata & a \\
13 24 39.14 & -32 06 33.4 & 14.90$\pm$0.10 & 16.42$\pm$0.05 & 1.52$\pm$0.11 &  0.63 & 0.59 & U & 0.09 & 2.4 & 13650 & \nodata \\
13 24 42.77 & -32 05 48.2 & 16.64$\pm$0.10 & 18.08$\pm$0.06 & 1.44$\pm$0.12 &  1.63 & 0.85 & R & 0.09 & 1.2 & \nodata & \nodata \\
13 24 47.57 & -31 20 29.9 & 14.47$\pm$0.10 & 15.84$\pm$0.05 & 1.37$\pm$0.11 &  1.66 & 0.90 & R & 0.07 & 2.9 &  9626 & \nodata \\
13 24 50.44 & -31 33 37.8 & 15.88$\pm$0.10 & 16.67$\pm$0.05 & 0.79$\pm$0.11 &  1.16 & 0.80 & R & 0.08 & 2.4 & 14773 & \nodata \\
13 24 57.54 & -30 51 01.1 & 16.64$\pm$0.10 & 17.73$\pm$0.06 & 1.09$\pm$0.12 &  0.58 & 0.70 & R & 0.13 & 3.5 & \nodata & \nodata \\
13 25 04.19 & -32 06 57.7 & 17.07$\pm$0.11 & 18.38$\pm$0.08 & 1.31$\pm$0.14 &  0.94 & 0.64 & R & 0.09 & 3.0 & \nodata & \nodata \\
13 25 07.47 & -32 05 19.0 & 17.05$\pm$0.10 & 18.68$\pm$0.09 & 1.63$\pm$0.13 &  0.37 & 0.56 & U & 0.09 & 2.9 & \nodata & \nodata \\
13 25 08.77 & -31 00 35.5 & 15.58$\pm$0.10 & 16.43$\pm$0.05 & 0.85$\pm$0.11 &  1.02 & 0.46 & R & 0.09 & 4.2 & \nodata & b \\
13 25 15.31 & -30 55 07.3 & 16.82$\pm$0.10 & 17.85$\pm$0.08 & 1.03$\pm$0.13 &  0.62 & 0.78 & R & 0.11 & 1.8 & \nodata & \nodata \\
13 25 29.81 & -31 15 59.8 & 15.54$\pm$0.10 & 16.71$\pm$0.06 & 1.17$\pm$0.12 &  0.46 & 0.40 & U & 0.08 & 1.7 & \nodata & \nodata \\
13 25 42.26 & -32 03 44.0 & 14.97$\pm$0.10 & 16.38$\pm$0.05 & 1.41$\pm$0.11 &  0.44 & 0.53 & R & 0.09 & 2.8 & \nodata & \nodata \\
13 25 50.32 & -32 08 54.2 & 13.95$\pm$0.10 & 15.64$\pm$0.05 & 1.69$\pm$0.11 &  0.62 & 0.64 & R & 0.09 & 4.1 & 13853 & \nodata \\
13 26 17.31 & -31 38 46.0 & 15.70$\pm$0.10 & 17.76$\pm$0.09 & 2.06$\pm$0.13 &  5.93 & 5.24 & R & 0.08 & 1.4 & 39812 & \nodata \\
13 26 18.81 & -30 57 37.0 & 14.06$\pm$0.10 & 15.61$\pm$0.05 & 1.55$\pm$0.11 &  5.31 & 4.60 & R & 0.09 & 2.5 & 14212 & \nodata \\
13 26 25.80 & -31 27 20.9 & 14.92$\pm$0.10 & 16.59$\pm$0.05 & 1.67$\pm$0.11 &  0.68 & 0.77 & U & 0.08 & 3.3 & 13749 & \nodata \\
13 26 28.67 & -30 44 44.4 & 14.22$\pm$0.10 & 15.80$\pm$0.05 & 1.58$\pm$0.11 &  4.62 & 2.57 & R & 0.14 & 3.8 & \nodata & b \\
13 26 35.69 & -30 59 36.4 & 16.26$\pm$0.10 & 16.88$\pm$0.05 & 0.62$\pm$0.11 &  0.54 & 0.46 & R & 0.08 & 3.8 & 14273 & \nodata \\
13 26 44.47 & -32 01 50.9 & 16.84$\pm$0.10 & 17.92$\pm$0.07 & 1.08$\pm$0.12 &  0.33 & 0.42 & U & 0.08 & 5.0 & \nodata & \nodata \\
13 26 48.60 & -31 34 18.6 & 16.02$\pm$0.10 & 16.81$\pm$0.06 & 0.79$\pm$0.12 &  2.15 & 1.34 & R & 0.08 & 2.7 & 14657 & \nodata \\
13 26 49.79 & -31 23 44.5 & 14.31$\pm$0.10 & 15.76$\pm$0.05 & 1.45$\pm$0.11 &  9.31 & 6.37 & R & 0.09 & 1.9 & 15190 & \nodata \\
13 26 49.94 & -30 37 56.9 & 16.05$\pm$0.10 & 17.54$\pm$0.07 & 1.49$\pm$0.12 &  1.91 & 0.64 & R & 0.24 & 3.5 & 14555 & \nodata \\
13 26 52.95 & -32 00 28.6 & 14.71$\pm$0.10 & 16.21$\pm$0.05 & 1.50$\pm$0.11 &  0.26 & 0.40 & U & 0.08 & 3.7 & 14389 & \nodata \\
13 26 53.88 & -31 18 55.9 & 17.20$\pm$0.11 & 18.52$\pm$0.13 & 1.32$\pm$0.17 &  1.40 & 1.08 & R & 0.10 & 1.1 & \nodata & d \\
13 26 54.01 & -31 18 39.5 & 16.56$\pm$0.10 & 18.61$\pm$0.14 & 2.05$\pm$0.17 &  0.46 & 0.50 & R & 0.10 & 2.6 & 14512 & d \\
13 26 57.51 & -32 11 52.3 & 15.56$\pm$0.10 & 16.38$\pm$0.05 & 0.82$\pm$0.11 &  0.78 & 0.55 & R & 0.10 & 4.6 & 13823 & \nodata \\
13 26 58.44 & -30 45 10.7 & 16.79$\pm$0.10 & 18.09$\pm$0.08 & 1.30$\pm$0.13 &  0.55 & 0.80 & U & 0.12 & 1.2 & \nodata & \nodata \\
13 27 03.35 & -31 13 31.5 & 14.98$\pm$0.10 & 16.32$\pm$0.05 & 1.34$\pm$0.11 &  1.13 & 0.51 & R & 0.08 & 2.7 & 16387 & \nodata \\
13 27 04.06 & -31 32 16.3 & 16.96$\pm$0.11 & 18.47$\pm$0.12 & 1.51$\pm$0.16 &  0.43 & 0.39 & U & 0.08 & 3.9 & \nodata & \nodata \\
13 27 29.66 & -31 23 24.3 & 13.69$\pm$0.10 & 15.55$\pm$0.05 & 1.86$\pm$0.11 &  3.47 & 2.93 & R & 0.09 & 0.6 & 14281 & \nodata \\
13 27 29.66 & -30 33 31.8 & 15.29$\pm$0.10 & 16.77$\pm$0.05 & 1.48$\pm$0.11 &  4.22 & 4.46 & R & 0.24 & 1.2 & 14150 & \nodata \\
13 27 34.96 & -31 32 59.2 & 14.57$\pm$0.10 & 16.27$\pm$0.05 & 1.70$\pm$0.11 &  1.06 & 0.79 & U & 0.09 & 1.9 & 12815 & \nodata \\
13 27 45.50 & -31 47 51.0 & 13.83$\pm$0.10 & 15.70$\pm$0.05 & 1.87$\pm$0.11 &  0.44 & 0.48 & U & 0.08 & 1.6 & 13186 & \nodata \\
13 27 51.48 & -31 14 45.1 & 16.74$\pm$0.10 & 17.36$\pm$0.06 & 0.62$\pm$0.12 &  1.56 & 1.29 & R & 0.08 & 2.7 & 14151 & \nodata \\
13 27 54.95 & -31 32 18.6 & 13.87$\pm$0.10 & 15.81$\pm$0.05 & 1.94$\pm$0.11 &  9.81 & 9.26 & R & 0.10 & 0.6 & 15481 & \nodata \\
13 27 56.84 & -31 29 44.6 & 13.58$\pm$0.10 & 15.46$\pm$0.05 & 1.88$\pm$0.11 &  5.51 & 4.60 & R & 0.10 & 1.7 & 14061 & \nodata \\
13 28 02.63 & -31 45 21.3 & 14.23$\pm$0.10 & 16.05$\pm$0.05 & 1.82$\pm$0.11 & 32.37 & 21.98 & R & 0.08 & 1.3 & 12857 & b,c,e \\
13 28 03.55 & -30 40 50.9 & 15.82$\pm$0.10 & 17.45$\pm$0.07 & 1.63$\pm$0.12 &  1.46 & 0.85 & R & 0.12 & 4.6 & 14996 & \nodata \\
13 28 08.02 & -32 08 46.1 & 14.99$\pm$0.10 & 16.40$\pm$0.05 & 1.41$\pm$0.11 &  2.38 & 1.58 & R & 0.08 & 3.5 & 21712 & \nodata \\
13 28 10.37 & -31 20 57.4 & 15.46$\pm$0.10 & 16.69$\pm$0.05 & 1.23$\pm$0.11 &  0.92 & 0.50 & R & 0.09 & 3.4 & 15690 & \nodata \\
13 28 10.85 & -32 12 51.6 & 16.42$\pm$0.10 & 18.40$\pm$0.07 & 1.98$\pm$0.12 &  0.62 & 0.36 & R & 0.10 & 6.8 & 43169 & \nodata \\
13 28 25.02 & -31 51 28.3 & 14.34$\pm$0.10 & 15.20$\pm$0.05 & 0.86$\pm$0.11 &  2.68 & 1.13 & R & 0.09 & 2.1 &  4389 & b \\
13 28 25.75 & -30 38 08.2 & 16.92$\pm$0.11 & 18.53$\pm$0.11 & 1.61$\pm$0.16 &  1.20 & 0.42 & R & 0.11 & 3.6 & \nodata & f \\
13 28 29.84 & -32 14 35.6 & 15.56$\pm$0.10 & 16.88$\pm$0.05 & 1.32$\pm$0.11 &  2.69 & 2.36 & R & 0.09 & 0.5 & \nodata & \nodata \\
13 28 30.11 & -31 47 15.7 & 15.74$\pm$0.10 & 16.66$\pm$0.05 & 0.92$\pm$0.11 &  0.52 & 0.45 & U & 0.09 & 4.5 &  9971 & \nodata \\
13 28 30.27 & -32 04 42.5 & 17.30$\pm$0.11 & 19.44$\pm$0.11 & 2.14$\pm$0.16 &  0.24 & 0.40 & U & 0.08 & 2.7 & \nodata & a \\
13 28 33.79 & -31 26 28.3 & 15.69$\pm$0.10 & 16.64$\pm$0.05 & 0.95$\pm$0.11 &  0.86 & 0.53 & R & 0.10 & 2.8 & 14522 & \nodata \\
13 28 34.83 & -31 44 44.4 & 15.30$\pm$0.10 & 16.20$\pm$0.05 & 0.90$\pm$0.11 &  0.62 & 0.43 & R & 0.10 & 0.9 &  9920 & \nodata \\
13 28 35.78 & -31 30 54.3 & 16.24$\pm$0.10 & 17.83$\pm$0.09 & 1.59$\pm$0.13 &  0.65 & 0.60 & R & 0.11 & 2.4 & 14003 & \nodata \\
13 28 38.47 & -32 10 36.7 & 16.42$\pm$0.10 & 17.17$\pm$0.05 & 0.75$\pm$0.11 &  0.65 & 0.42 & U & 0.08 & 4.0 & \nodata & \nodata \\
13 28 38.72 & -31 20 49.3 & 14.28$\pm$0.10 & 16.06$\pm$0.05 & 1.78$\pm$0.11 &  0.61 & 0.41 & R & 0.10 & 0.9 & 15625 & \nodata \\
13 28 39.26 & -31 39 41.1 & 16.46$\pm$0.10 & 17.16$\pm$0.06 & 0.70$\pm$0.12 &  0.83 & 0.96 & U & 0.11 & 2.7 & 13314 & \nodata \\
13 28 50.26 & -30 47 23.1 & 17.22$\pm$0.11 & 18.84$\pm$0.15 & 1.62$\pm$0.19 &  0.67 & 0.52 & R & 0.08 & 2.3 & \nodata & \nodata \\
13 28 50.91 & -32 07 53.1 & 14.96$\pm$0.10 & 16.02$\pm$0.05 & 1.06$\pm$0.11 &  2.00 & 1.56 & R & 0.08 & 2.4 & 15073 & \nodata \\
13 28 52.29 & -32 09 45.8 & 14.99$\pm$0.10 & 16.83$\pm$0.05 & 1.84$\pm$0.11 &  7.29 & 6.05 & R & 0.08 & 1.8 & 12441 & \nodata \\
13 28 54.30 & -31 55 11.1 & 15.60$\pm$0.10 & 16.77$\pm$0.05 & 1.17$\pm$0.11 &  0.77 & 0.70 & U & 0.08 & 1.9 & 11937 & \nodata \\
13 28 57.71 & -31 51 34.9 & 15.50$\pm$0.10 & 17.04$\pm$0.05 & 1.54$\pm$0.11 &  1.60 & 1.78 & U & 0.09 & 1.2 & 16532 & g \\
13 28 59.29 & -31 44 15.7 & 16.18$\pm$0.10 & 17.60$\pm$0.09 & 1.42$\pm$0.13 &  1.58 & 1.20 & R & 0.10 & 2.7 & 13550 & \nodata \\
13 29 04.33 & -31 47 52.4 & 15.83$\pm$0.10 & 16.62$\pm$0.05 & 0.79$\pm$0.11 &  0.69 & 0.67 & R & 0.09 & 0.8 & 15411 & \nodata \\
13 29 12.29 & -31 18 39.9 & 15.37$\pm$0.10 & 17.00$\pm$0.06 & 1.63$\pm$0.12 &  0.55 & 0.35 & R & 0.10 & 1.0 & 13610 & \nodata \\
13 29 14.98 & -31 16 44.9 & 14.89$\pm$0.10 & 16.24$\pm$0.05 & 1.35$\pm$0.11 &  1.07 & 0.86 & R & 0.10 & 3.5 & 13573 & b \\
13 29 17.00 & -32 02 33.9 & 14.46$\pm$0.10 & 16.36$\pm$0.06 & 1.90$\pm$0.12 &  1.13 & 0.85 & R & 0.08 & 2.7 & 13834 & \nodata \\
13 29 20.43 & -30 46 37.4 & 15.56$\pm$0.10 & 17.22$\pm$0.06 & 1.66$\pm$0.12 &  1.62 & 1.51 & R & 0.09 & 1.2 & 14398 & \nodata \\
13 29 28.57 & -31 39 25.7 & 15.99$\pm$0.10 & 16.80$\pm$0.06 & 0.81$\pm$0.12 &  2.30 & 1.81 & R & 0.12 & 1.5 & 14413 & \nodata \\
13 29 31.44 & -31 37 29.0 & 16.03$\pm$0.10 & 17.46$\pm$0.06 & 1.43$\pm$0.12 &  0.88 & 0.70 & R & 0.11 & 3.1 & 15135 & \nodata \\
13 29 34.78 & -30 32 43.6 & 14.45$\pm$0.10 & 15.74$\pm$0.05 & 1.29$\pm$0.11 &  1.58 & 1.27 & R & 0.14 & 3.7 &  4575 & \nodata \\
13 29 37.24 & -30 57 56.5 & 14.57$\pm$0.10 & 16.30$\pm$0.05 & 1.73$\pm$0.11 &  0.83 & 0.49 & R & 0.10 & 2.2 & 10428 & \nodata \\
13 29 43.22 & -30 28 22.9 & 15.82$\pm$0.10 & 17.13$\pm$0.06 & 1.31$\pm$0.12 &  1.31 & 0.87 & R & 0.25 & 5.5 & \nodata & \nodata \\
13 29 50.90 & -31 22 58.2 & 17.02$\pm$0.10 & 18.94$\pm$0.17 & 1.92$\pm$0.20 & 16.72 & 6.12 & R & 0.11 & 0.5 & 58775 & b,c \\
13 29 54.12 & -31 08 43.9 & 15.76$\pm$0.10 & 17.25$\pm$0.06 & 1.49$\pm$0.12 &  0.56 & 0.33 & U & 0.10 & 0.5 & \nodata & \nodata \\
13 29 54.20 & -32 00 59.3 & 15.32$\pm$0.10 & 17.13$\pm$0.06 & 1.81$\pm$0.12 &  1.03 & 0.78 & R & 0.08 & 0.9 & 14012 & \nodata \\
13 29 55.10 & -31 19 56.4 & 14.38$\pm$0.10 & 16.11$\pm$0.05 & 1.73$\pm$0.11 &  3.87 & 1.54 & R & 0.11 & 2.9 & 14815 & \nodata \\
13 29 57.07 & -31 11 08.1 & 15.40$\pm$0.10 & 16.95$\pm$0.06 & 1.55$\pm$0.12 &  1.97 & 1.84 & R & 0.10 & 0.9 & 16628 & \nodata \\
13 29 59.97 & -32 25 59.2 & 16.95$\pm$0.11 & 20.20$\pm$0.47 & 3.25$\pm$0.48 &  0.63 & 1.57 & R & 0.11 & 4.2 & \nodata & a \\
13 30 05.14 & -31 24 36.8 & 14.73$\pm$0.10 & 16.43$\pm$0.05 & 1.70$\pm$0.11 &  8.40 & 2.67 & R & 0.12 & 3.7 & 13780 & \nodata \\
13 30 05.20 & -31 13 53.4 & 16.17$\pm$0.10 & 17.18$\pm$0.06 & 1.01$\pm$0.11 &  0.95 & 0.84 & R & 0.11 & 1.2 & 15943 & \nodata \\
13 30 22.06 & -30 43 52.1 & 15.31$\pm$0.10 & 17.15$\pm$0.06 & 1.84$\pm$0.12 &  1.71 & 0.94 & R & 0.13 & 2.3 & \nodata & \nodata \\
13 30 24.83 & -30 41 37.1 & 14.94$\pm$0.10 & 16.34$\pm$0.05 & 1.40$\pm$0.11 &  1.33 & 1.24 & R & 0.14 & 2.8 & 14491 & \nodata \\
13 30 32.80 & -31 34 33.8 & 13.98$\pm$0.10 & 15.71$\pm$0.05 & 1.73$\pm$0.11 &  1.48 & 0.63 & R & 0.12 & 0.5 & 14478 & b\\
13 30 35.76 & -30 41 07.8 & 15.75$\pm$0.10 & 17.44$\pm$0.07 & 1.69$\pm$0.12 &  1.25 & 1.18 & R & 0.16 & 6.3 & \nodata & \nodata \\
13 30 39.21 & -30 48 08.7 & 16.49$\pm$0.10 & 17.71$\pm$0.07 & 1.22$\pm$0.12 &  0.59 & 0.53 & R & 0.11 & 5.6 & \nodata & \nodata \\
13 30 40.65 & -31 42 32.1 & 17.03$\pm$0.10 & 18.37$\pm$0.10 & 1.34$\pm$0.14 &  0.53 & 0.34 & R & 0.09 & 1.9 & 15489 & \nodata \\
13 30 40.74 & -31 51 56.0 & 15.65$\pm$0.10 & 17.28$\pm$0.06 & 1.63$\pm$0.12 &  0.43 & 0.35 & R & 0.08 & 3.5 & 14407 & \nodata \\
13 30 41.47 & -31 34 17.3 & 15.19$\pm$0.10 & 16.67$\pm$0.06 & 1.48$\pm$0.12 &  1.56 & 1.33 & R & 0.10 & 6.1 & 15323 & d \\
13 30 42.03 & -31 34 28.7 & 14.15$\pm$0.10 & 15.24$\pm$0.05 & 1.09$\pm$0.11 &  1.69 & 0.97 & R & 0.10 & 3.0 & 15215 & d,h \\
13 30 46.20 & -32 18 42.7 & 16.85$\pm$0.10 & 18.29$\pm$0.09 & 1.44$\pm$0.13 &  0.93 & 0.68 & R & 0.10 & 1.8 & \nodata & \nodata \\
13 30 47.73 & -31 51 06.5 & 14.95$\pm$0.10 & 16.53$\pm$0.05 & 1.58$\pm$0.11 &  0.55 & 0.40 & R & 0.08 & 1.7 & 14636 & \nodata \\
13 30 48.93 & -31 43 25.7 & 12.64$\pm$0.10 & 13.75$\pm$0.05 & 1.11$\pm$0.11 &  7.53 & 1.42 & R & 0.09 & 1.3 & 13226 & b,c,i \\
13 30 51.04 & -31 44 17.9 & 14.90$\pm$0.10 & 16.66$\pm$0.06 & 1.76$\pm$0.12 &  0.67 & 0.57 & U & 0.09 & 3.4 & 13815 & \nodata \\
13 30 51.82 & -31 04 40.3 & 15.83$\pm$0.10 & 17.16$\pm$0.06 & 1.33$\pm$0.12 &  3.78 & 2.79 & R & 0.10 & 1.6 & \nodata & \nodata \\
13 30 52.20 & -31 46 07.0 & 16.07$\pm$0.10 & 17.45$\pm$0.06 & 1.38$\pm$0.12 &  1.68 & 1.68 & R & 0.08 & 1.1 & 13832 & \nodata \\
13 30 52.70 & -31 27 13.2 & 15.74$\pm$0.10 & 17.27$\pm$0.06 & 1.53$\pm$0.12 &  0.78 & 0.95 & U & 0.11 & 1.8 & 14612 & \nodata \\
13 30 53.83 & -31 30 21.7 & 15.32$\pm$0.10 & 16.66$\pm$0.05 & 1.34$\pm$0.11 &  0.50 & 0.40 & R & 0.10 & 3.5 & 14409 & \nodata \\
13 30 56.16 & -31 33 42.9 & 16.49$\pm$0.10 & 17.75$\pm$0.07 & 1.26$\pm$0.12 &  0.56 & 0.39 & R & 0.09 & 3.9 & 14348 & \nodata \\
13 31 00.75 & -31 44 53.9 & 15.55$\pm$0.10 & 16.73$\pm$0.05 & 1.18$\pm$0.11 &  2.36 & 2.03 & R & 0.08 & 0.4 & 11976 & \nodata \\
13 31 01.99 & -31 40 36.3 & 16.61$\pm$0.10 & 17.55$\pm$0.07 & 0.94$\pm$0.12 &  0.76 & 0.72 & U & 0.09 & 1.9 & 13180 & \nodata \\
13 31 16.90 & -31 35 38.4 & 15.66$\pm$0.10 & 16.71$\pm$0.05 & 1.05$\pm$0.11 &  1.96 & 0.86 & R & 0.09 & 0.6 & 14578 & b \\
13 31 17.43 & -31 24 34.7 & 15.66$\pm$0.10 & 17.54$\pm$0.06 & 1.88$\pm$0.12 &  1.40 & 1.01 & R & 0.09 & 4.3 & 13763 & \nodata \\
13 31 19.47 & -31 13 44.4 & 14.65$\pm$0.10 & 16.29$\pm$0.05 & 1.64$\pm$0.11 &  3.79 & 1.99 & R & 0.09 & 0.7 & 15710 & b \\
13 31 20.22 & -31 44 34.6 & 15.70$\pm$0.10 & 16.89$\pm$0.06 & 1.19$\pm$0.12 &  2.37 & 1.16 & R & 0.08 & 3.4 & 13333 & \nodata \\
13 31 22.42 & -31 22 10.4 & 16.77$\pm$0.10 & 17.68$\pm$0.07 & 0.91$\pm$0.12 &  0.52 & 0.43 & R & 0.09 & 2.1 & 14027 & \nodata \\
13 31 27.58 & -31 49 14.7 & 14.23$\pm$0.10 & 16.13$\pm$0.05 & 1.90$\pm$0.11 &  1.64 & 1.59 & U & 0.08 & 1.9 & 12928 & \nodata \\
13 31 37.79 & -31 33 12.0 & 14.97$\pm$0.10 & 16.14$\pm$0.05 & 1.17$\pm$0.11 &  2.92 & 2.28 & R & 0.08 & 0.8 & 14194 & j \\
13 31 43.87 & -30 55 44.3 & 15.65$\pm$0.10 & 16.93$\pm$0.05 & 1.28$\pm$0.11 &  1.25 & 0.73 & R & 0.12 & 3.5 & \nodata & f \\
13 31 53.37 & -31 55 36.1 & 17.14$\pm$0.11 & 18.63$\pm$0.06 & 1.49$\pm$0.13 &  0.58 & 0.72 & U & 0.07 & 0.8 & \nodata & \nodata \\
13 31 54.87 & -31 54 04.1 & 15.72$\pm$0.10 & 16.79$\pm$0.05 & 1.07$\pm$0.11 &  1.94 & 1.13 & R & 0.07 & 1.9 & 13043 & \nodata \\
13 31 59.81 & -31 49 22.2 & 16.53$\pm$0.10 & 16.91$\pm$0.05 & 0.38$\pm$0.11 &  0.89 & 0.89 & U & 0.07 & 2.7 & 12963 & \nodata \\
13 32 00.41 & -32 24 01.8 & 16.34$\pm$0.10 & 17.63$\pm$0.06 & 1.29$\pm$0.12 &  1.20 & 0.99 & R & 0.12 & 5.8 & \nodata & k \\
13 32 01.52 & -31 13 58.0 & 14.75$\pm$0.10 & 15.99$\pm$0.05 & 1.24$\pm$0.11 &  1.68 & 1.77 & U & 0.08 & 2.4 & \nodata & \nodata \\
13 32 03.17 & -31 46 48.6 & 13.99$\pm$0.10 & 15.65$\pm$0.05 & 1.66$\pm$0.11 &  4.30 & 3.74 & R & 0.08 & 1.0 & 13107 & \nodata \\
13 32 12.62 & -31 05 42.8 & 16.46$\pm$0.10 & 17.79$\pm$0.06 & 1.33$\pm$0.12 &  0.52 & 0.47 & R & 0.07 & 0.8 & \nodata & \nodata \\
13 32 15.15 & -31 13 29.8 & 17.28$\pm$0.12 & 18.78$\pm$0.07 & 1.50$\pm$0.14 &  0.76 & 0.49 & U & 0.07 & 4.0 & 44703 & \nodata \\
13 32 20.74 & -32 07 37.6 & 16.66$\pm$0.11 & 18.52$\pm$0.07 & 1.86$\pm$0.13 &  1.27 & 1.05 & R & 0.09 & 1.5 & \nodata & a \\
13 32 30.07 & -31 03 15.8 & 16.13$\pm$0.10 & 17.44$\pm$0.05 & 1.31$\pm$0.11 &  3.07 & 2.47 & R & 0.08 & 0.6 & 13772 & \nodata \\
13 32 35.44 & -31 31 45.5 & 16.19$\pm$0.10 & 16.74$\pm$0.05 & 0.55$\pm$0.11 &  1.80 & 1.20 & R & 0.07 & 3.3 & 15193 & \nodata \\
13 32 36.13 & -32 20 01.3 & 16.55$\pm$0.10 & 19.03$\pm$0.14 & 2.48$\pm$0.17 &  2.22 & 1.61 & R & 0.10 & 1.6 & \nodata & a \\
13 32 38.27 & -31 01 41.3 & 16.36$\pm$0.10 & 17.40$\pm$0.05 & 1.04$\pm$0.11 &  1.19 & 1.34 & U & 0.09 & 1.1 & 13189 & \nodata \\
13 32 42.95 & -31 55 13.7 & 15.86$\pm$0.10 & 16.84$\pm$0.05 & 0.98$\pm$0.11 &  0.42 & 0.33 & U & 0.08 & 4.1 & 14483 & \nodata \\
13 32 43.89 & -31 47 01.3 & 15.69$\pm$0.10 & 16.71$\pm$0.05 & 1.02$\pm$0.11 &  0.76 & 0.68 & U & 0.07 & 1.0 & 12984 & \nodata \\
13 32 46.48 & -31 37 03.3 & 15.38$\pm$0.10 & 16.33$\pm$0.05 & 0.95$\pm$0.11 &  1.83 & 1.37 & R & 0.08 & 2.2 & 12315 & \nodata \\
13 32 46.75 & -31 57 59.2 & 14.20$\pm$0.10 & 16.07$\pm$0.05 & 1.87$\pm$0.11 &  2.22 & 1.76 & R & 0.08 & 1.5 & 11106 & \nodata \\
13 33 00.85 & -30 53 07.4 & 15.22$\pm$0.10 & 16.67$\pm$0.05 & 1.45$\pm$0.11 &  2.05 & 1.35 & R & 0.16 & 2.8 & 10561 & \nodata \\
13 33 01.00 & -31 13 29.3 & 16.00$\pm$0.10 & 17.44$\pm$0.06 & 1.44$\pm$0.12 &  1.22 & 1.18 & R & 0.07 & 1.2 & 10981 & \nodata \\
13 33 01.33 & -31 44 19.2 & 17.30$\pm$0.10 & 18.80$\pm$0.18 & 1.50$\pm$0.21 &  1.69 & 1.63 & R & 0.07 & 2.2 & \nodata & \nodata \\
13 33 04.16 & -31 39 04.0 & 16.24$\pm$0.10 & 17.69$\pm$0.08 & 1.45$\pm$0.13 &  2.06 & 1.55 & R & 0.07 & 0.1 & 14120 & \nodata \\
13 33 14.20 & -32 12 08.2 & 16.51$\pm$0.10 & 18.62$\pm$0.12 & 2.11$\pm$0.16 &  4.55 & 1.96 & R & 0.08 & 2.6 & \nodata & a \\
13 33 17.99 & -31 39 11.1 & 16.7\phn$\pm$0.2\phn & 17.5\phn$\pm$0.1\phn & 0.8\phn$\pm$0.2\phn & 0.93 & 0.77 & R & 0.08 & 2.7 & \nodata & h,l \\
13 33 23.53 & -31 24 23.3 & 15.50$\pm$0.10 & 16.80$\pm$0.05 & 1.30$\pm$0.11 &  0.25 &  0.39 & U & 0.07 & 1.8 & 13103 & \nodata \\
13 33 27.77 & -31 02 16.2 & 17.35$\pm$0.11 & 20.4\phn$\pm$0.5\phn & 3.1\phn$\pm$0.5\phn & 0.66 & 0.77 & U & 0.11 & 2.0 & \nodata & a,m \\
13 33 28.21 & -32 01 07.1 & 15.61$\pm$0.10 & 16.79$\pm$0.05 & 1.18$\pm$0.11 & 0.85 & 0.91 & R & 0.07 & 0.5 & \nodata & \nodata \\
13 33 31.63 & -31 40 58.6 & 15.39$\pm$0.10 & 17.33$\pm$0.06 & 1.94$\pm$0.12 & 103.34 & 35.58 & R & 0.09 & 9.6 & 14438 & b,c \\
13 33 34.76 & -31 40 20.5 & 14.04$\pm$0.10 & 15.92$\pm$0.05 & 1.88$\pm$0.11 & 0.51 & 0.40 & U & 0.09 & 0.7 & 14653 & \nodata \\
13 33 40.7\phn & -31 19 57.\phn & 15.8\phn$\pm$0.2\phn & 17.0\phn$\pm$0.2\phn & 1.2\phn$\pm$0.3\phn & 6.25 & 4.82 & R & 0.08 & 1.0 & \nodata & n \\
13 33 40.94 & -31 54 23.4 & 16.43$\pm$0.10 & 17.37$\pm$0.06 & 0.94$\pm$0.12 & 0.71 & 0.62 & R & 0.07 & 3.0 & \nodata & \nodata \\
13 33 50.43 & -31 28 38.0 & 15.43$\pm$0.10 & 16.40$\pm$0.05 & 0.97$\pm$0.11 & 0.67 & 0.61 & U & 0.08 & 3.0 & 10994 & \nodata \\
13 33 57.54 & -31 29 05.5 & 16.05$\pm$0.10 & 16.94$\pm$0.06 & 0.89$\pm$0.12 & 1.17 & 0.78 & R & 0.08 & 2.2 & \nodata & o \\
13 33 58.73 & -31 58 16.1 & 15.93$\pm$0.10 & 17.23$\pm$0.05 & 1.30$\pm$0.11 & 1.29 & 1.20 & U & 0.07 & 0.4 & 14689 & \nodata \\
13 34 00.12 & -32 22 56.1 & 16.54$\pm$0.10 & 17.74$\pm$0.07 & 1.20$\pm$0.12 & 1.36 & 0.91 & R & 0.11 & 3.7 & \nodata & \nodata \\
13 34 06.6\phn & -31 43 39.\phn & 16.4\phn$\pm$0.2\phn & 17.4\phn$\pm$0.2\phn & 1.0\phn$\pm$0.3\phn & 1.05 & 0.86 & R & 0.08 & 2.6 & \nodata & n \\
13 34 08.12 & -31 16 55.6 & 16.91$\pm$0.10 & 18.17$\pm$0.09 & 1.26$\pm$0.13 & 0.90 & 0.45 & R & 0.07 & 1.8 & \nodata & \nodata \\
13 34 18.21 & -31 49 08.1 & 16.02$\pm$0.10 & 17.53$\pm$0.08 & 1.51$\pm$0.13 & 0.71 & 0.72 & R & 0.07 & 3.0 & \nodata & \nodata \\
13 34 19.43 & -31 59 14.4 & 15.31$\pm$0.10 & 16.77$\pm$0.05 & 1.46$\pm$0.11 & 0.28 & 0.40 & U & 0.07 & 2.3 & 10908 & \nodata \\
13 34 29.81 & -31 36 55.2 & 16.70$\pm$0.10 & 18.20$\pm$0.09 & 1.50$\pm$0.13 & 0.40 & 0.34 & U & 0.06 & 2.6 & \nodata & \nodata \\
13 34 31.48 & -32 05 09.0 & 15.54$\pm$0.10 & 16.64$\pm$0.05 & 1.10$\pm$0.11 & 1.00 & 0.59 & R & 0.07 & 2.3 & 12647 & \nodata \\
13 34 36.19 & -31 41 06.7 & 13.81$\pm$0.10 & 14.89$\pm$0.05 & 1.08$\pm$0.11 & 1.47 & 1.23 & R & 0.07 & 5.4 & 11407 & \nodata \\
13 34 39.94 & -31 32 54.5 & 15.91$\pm$0.10 & 17.34$\pm$0.06 & 1.43$\pm$0.12 & 5.37 & 3.99 & R & 0.06 & 6.4 & 16490 & \nodata \\
13 34 40.45 & -31 29 43.2 & 16.42$\pm$0.10 & 17.92$\pm$0.08 & 1.50$\pm$0.13 & 0.66 & 0.61 & R & 0.06 & 3.8 & \nodata & \nodata \\
13 34 40.80 & -31 26 43.6 & 15.06$\pm$0.10 & 16.36$\pm$0.05 & 1.30$\pm$0.11 & 1.57 & 1.55 & R & 0.06 & 1.0 & 14156 & \nodata \\
13 34 43.00 & -32 22 14.0 & 14.97$\pm$0.10 & 16.11$\pm$0.05 & 1.14$\pm$0.11 & 1.12 & 0.67 & R & 0.09 & 1.5 & 12322 & \nodata \\
13 34 43.81 & -31 45 52.2 & 15.86$\pm$0.10 & 17.03$\pm$0.06 & 1.17$\pm$0.12 & 0.29 & 0.41 & U & 0.08 & 1.9 & \nodata & \nodata \\
13 34 57.25 & -32 26 31.6 & 16.61$\pm$0.10 & 18.27$\pm$0.08 & 1.66$\pm$0.13 & 0.78 & 0.85 & R & 0.11 & 0.6 & \nodata & \nodata \\
13 35 03.06 & -31 39 18.6 & 14.00$\pm$0.10 & 15.85$\pm$0.05 & 1.85$\pm$0.11 & 15.49 & 14.40 & R & 0.07 & 1.1 & 15205 & \nodata \\
13 35 12.61 & -31 33 45.0 & 15.22$\pm$0.10 & 16.33$\pm$0.05 & 1.11$\pm$0.11 & 2.33 & 1.37 & R & 0.07 & 1.4 & 13673 & \nodata \\
13 35 15.57 & -31 30 48.2 & 16.80$\pm$0.10 & 17.98$\pm$0.08 & 1.18$\pm$0.13 & 0.59 & 0.39 & U & 0.06 & 5.7 & \nodata & \nodata \\
13 35 18.70 & -32 32 30.4 & 15.08$\pm$0.10 & 16.62$\pm$0.05 & 1.54$\pm$0.11 & 4.71 & 2.85 & R & 0.15 & 4.9 & \nodata & \nodata \\
13 35 18.90 & -31 34 57.5 & 16.28$\pm$0.10 & 19.70$\pm$0.27 & 3.42$\pm$0.29 & 0.71 & 0.81 & U & 0.07 & 1.3 & 58050 & \nodata \\
13 35 29.53 & -30 52 56.1 & 14.88$\pm$0.10 & 15.98$\pm$0.05 & 1.10$\pm$0.11 & 2.07 & 2.46 & U & 0.21 & 1.5 & 12117 & \nodata \\
13 35 32.44 & -31 07 56.2 & 17.02$\pm$0.10 & 19.83$\pm$0.31 & 2.81$\pm$0.33 & 2.45 & 2.97 & U & 0.08 & 2.2 & \nodata & a \\
13 35 42.55 & -31 53 54.6 & 14.45$\pm$0.10 & 16.26$\pm$0.05 & 1.81$\pm$0.11 & 14.41 & 4.18 & R & 0.08 & 3.1 & 14385 & b,c \\
13 35 47.72 & -32 27 23.9 & 16.54$\pm$0.10 & 18.97$\pm$0.18 & 2.43$\pm$0.21 & 2.07 & 1.16 & R & 0.15 & 3.8 & \nodata & a \\
13 35 53.20 & -31 29 41.4 & 16.84$\pm$0.10 & 17.82$\pm$0.07 & 0.98$\pm$0.12 & 0.40 & 0.34 & R & 0.07 & 4.6 & \nodata & \nodata \\
13 35 56.85 & -32 01 50.0 & 15.96$\pm$0.10 & 18.29$\pm$0.11 & 2.33$\pm$0.15 & 1.99 & 0.83 & R & 0.09 & 0.6 & \nodata & a \\
13 35 59.75 & -31 23 04.6 & 16.08$\pm$0.10 & 17.59$\pm$0.07 & 1.51$\pm$0.12 & 0.26 & 0.39 & U & 0.07 & 6.1 & \nodata & \nodata \\
13 36 18.04 & -31 46 45.3 & 14.26$\pm$0.10 & 15.73$\pm$0.05 & 1.47$\pm$0.11 & 2.58 & 2.12 & R & 0.08 & 0.7 & 12263 & \nodata \\
13 36 18.70 & -31 48 31.2 & 15.01$\pm$0.10 & 16.15$\pm$0.05 & 1.14$\pm$0.11 & 2.76 & 2.32 & R & 0.07 & 0.7 & 11692 & \nodata \\
13 36 20.97 & -31 34 39.7 & 17.09$\pm$0.11 & 18.66$\pm$0.16 & 1.57$\pm$0.19 & 1.00 & 0.95 & U & 0.08 & 3.7 & \nodata & \nodata \\
13 36 26.46 & -31 06 36.4 & 17.15$\pm$0.10 & 18.69$\pm$0.13 & 1.54$\pm$0.16 & 0.80 & 0.84 & R & 0.12 & 1.0 & \nodata & \nodata \\
\enddata

\tablenotetext{a}{Probable background galaxy on the basis of color-magnitude relation.}
\tablenotetext{b}{Extended source. Flux calculated using TVSTAT.}
\tablenotetext{c}{See Figures \ref{fig-132145} -- \ref{fig-133542}.}
\tablenotetext{d}{These two galaxies are within the same radio emission, which has been deblended using JMFIT to produce the two detections. In the case of J132653-311838, its red color is inconsistent with its NED velocity.}
\tablenotetext{e}{Reported flux includes the possible ``tail'' to the radio galaxy. Without this extended component, the flux is about 25.7 mJy.}
\tablenotetext{f}{Source next to a bright source which is likely unrelated. The quoted radio data were determined using JMFIT to deblend the emission from the noted galaxy from the brighter source.}
\tablenotetext{g}{Part of galaxy pair which are blended in the present photometry. According to NED, the other galaxy in the pair is at 13:28:57.9 -31:51:36 and has a velocity of 16532 \kms, which has also been adopted for the radio galaxy.}
\tablenotetext{h}{Optical magnitude corrupted by foreground star.}
\tablenotetext{i}{See Section \ref{ssec-evolution} for a discussion of this galaxy.}
\tablenotetext{j}{Part of a galaxy pair. The other galaxy is at 13:31:37.56 -31:33:06.5 and has a velocity of 14194 \kms, which has been adopted for the radio galaxy.}
\tablenotetext{k}{Part of galaxy pair which are blended in the present photometry.}
\tablenotetext{l}{SExtractor only identified the galaxy in the $B$ image. The position provided is based on this information, with the $R$ magnitude determined manually (see text).}
\tablenotetext{m}{SExtractor only identified the galaxy in the $R$ image. The $B$ magnitude was determined manually (see text).}
\tablenotetext{n}{Not found by SExtractor due to nearby star. The $B$ and $R$ magnitudes were determined manually (see text).}
\tablenotetext{o}{Visually this galaxy appears to be part of a compact group of four galaxies. While NED contains no velocity for this particular galaxy, the brightest galaxy of the group (located at 13:33:58.33 -31:29:25.7) is a cluster elliptical with a velocity of 14399.}

\tablecomments{The listed RA and Dec correspond to the fitted centers of the galaxies in the optical (J2000). The radio data are the integral flux density ($S_I$) in mJy, the peak flux density ($S_P$) in mJy beam$^{-1}$, whether the source was resolved (R) or unresolved (U), and the local noise (rms) in mJy beam$^{-1}$ evaluated over a 6\arcmin{} box centered on the source. ``Sep'' refers to the separation in arcseconds between the peak of the radio emission and the optical position (if not provided, the radio source is extended and may be seen in Figures \ref{fig-132145} -- \ref{fig-133542}). All velocities are taken from NED and are in units of km s$^{-1}$.}

\end{deluxetable}

\begin{deluxetable}{l r r r}
\tablecolumns{4}
\tablecaption{Cluster Red Sequence Fits\label{tbl-rsfit}}
\tablewidth{0pt}
\tablehead{
\colhead{Cluster} & \colhead{Intercept} & \colhead{Slope} & \colhead{$N_{gal}$}
}
\startdata
A3556      & 2.86 & -0.080 &  317 \\
A3558      & 2.69 & -0.067 &  542 \\
A3562      & 2.79 & -0.072 &  378 \\
SC1327-312 & 3.05 & -0.092 &  585 \\
SC1329-313 & 3.15 & -0.098 &  488 \\
All data   & 2.49 & -0.054 & 1696 \\
\enddata

\tablecomments{The red sequence is parametrized as $B - R = \mbox{Intercept} - \mbox{Slope} \times R$. $N_{gal}$ refers to the total number of galaxies, after 3$\sigma$ clipping, which went into the fit. Except in the case of ``All data,'' a 2 Mpc radius has been adopted.}
\end{deluxetable}

\begin{deluxetable}{r r r c r r r r r c}
\tablecolumns{10}
\tablecaption{Absolute Quantities for Cluster Radio Galaxies\label{tbl-abs}}
\tabletypesize{\footnotesize}
\tablewidth{0pt}
\tablehead{
\colhead{RA} & \colhead{Dec} & \colhead{$M_R$} & \colhead{$L_{1.4GHz}$} & 
\colhead{$r_{A3556}$} & \colhead{$r_{A3558}$} & \colhead{$r_{A3562}$} & 
\colhead{$r_{SC1327}$} & \colhead{$r_{SC1329}$} & \colhead{Flags} \\
\colhead{(J2000)} & \colhead{(J2000)} & \colhead{} & \colhead{[$\times 10^{22}$ W Hz$^{-1}$]} &
\colhead{[Mpc]} & \colhead{[Mpc]} & \colhead{[Mpc]} & \colhead{[Mpc]} & \colhead{[Mpc]} &
\colhead{}
}
\startdata
\cutinhead{Galaxies in Radio Galaxy Fraction Analysis:}
13 22 06.75 & -31 46 16.4 & -22.1 & \phn3.02 & 1.30 & 3.75 & 7.21 & 4.86 & 5.99 & \nodata \\
13 23 29.07 & -31 50 39.6 & -22.5 & \phn1.75 & 0.67 & 2.98 & 6.35 & 4.03 & 5.12 & \nodata \\
13 23 54.22 & -32 10 56.3 & -22.0 & \phn2.37 & 1.55 & 3.25 & 6.23 & 4.07 & 4.96 & \nodata \\
13 23 57.58 & -31 38 45.2 & -22.4 & 22.9\phn & 0.10 & 2.54 & 6.05 & 3.68 & 4.85 & \nodata \\
13 24 06.68 & -31 40 12.0 & -22.9 & \phn4.28 & 0.03 & 2.46 & 5.95 & 3.59 & 4.75 & \nodata \\
13 24 22.77 & -31 42 23.7 & -22.5 & \phn5.72 & 0.22 & 2.32 & 5.78 & 3.42 & 4.57 & \nodata \\
13 24 32.46 & -31 06 58.2 & -20.7 & \phn1.03 & 1.64 & 2.42 & 5.95 & 3.64 & 4.94 & a \\
13 26 18.81 & -30 57 37.0 & -22.3 & \phn2.19 & 2.51 & 1.88 & 5.06 & 2.93 & 4.21 & \nodata \\
13 26 48.60 & -31 34 18.6 & -20.3 & \phn0.89 & 1.73 & 0.74 & 4.26 & 1.88 & 3.11 & \nodata \\
13 26 49.79 & -31 23 44.5 & -22.1 & \phn3.83 & 1.90 & 0.74 & 4.32 & 1.97 & 3.27 & \nodata \\
13 27 29.66 & -31 23 24.3 & -22.7 & \phn1.43 & 2.29 & 0.40 & 3.91 & 1.59 & 2.89 & \nodata \\
13 27 54.95 & -31 32 18.6 & -22.5 & \phn4.04 & 2.43 & 0.14 & 3.57 & 1.20 & 2.47 & \nodata \\
13 27 56.84 & -31 29 44.6 & -22.8 & \phn2.27 & 2.48 & 0.02 & 3.57 & 1.21 & 2.50 & \nodata \\
13 28 02.63 & -31 45 21.3 & -22.1 & 13.3\phn & 2.50 & 0.79 & 3.47 & 1.18 & 2.25 & \nodata \\
13 28 50.91 & -32 07 53.1 & -21.4 & \phn0.82 & 3.29 & 1.99 & 3.24 & 1.66 & 1.97 & \nodata \\
13 28 52.29 & -32 09 45.8 & -21.4 & \phn3.00 & 3.35 & 2.08 & 3.27 & 1.74 & 2.00 & b \\
13 28 57.71 & -31 51 34.9 & -20.9 & \phn0.73 & 3.12 & 1.28 & 2.93 & 0.91 & 1.67 & c \\
13 29 28.57 & -31 39 25.7 & -20.4 & \phn0.95 & 3.39 & 1.10 & 2.56 & 0.24 & 1.42 & \nodata \\
13 29 55.10 & -31 19 56.4 & {\it -22.0} & \phn1.59 & 3.81 & 1.36 & 2.50 & 0.82 & 1.78 & \nodata \\
13 29 57.07 & -31 11 08.1 & {\it -21.0} & \phn0.81 & 3.97 & 1.58 & 2.69 & 1.26 & 2.14 & \nodata \\
13 30 05.14 & -31 24 36.8 & -21.6 & \phn3.46 & 3.86 & 1.40 & 2.32 & 0.62 & 1.53 & \nodata \\
13 30 42.03 & -31 34 28.7 & -22.2 & \phn0.70 & 4.18 & 1.78 & 1.81 & 0.59 & 0.91 & \nodata \\
13 30 48.93 & -31 43 25.7 & -23.7 & \phn3.10 & 4.24 & 1.96 & 1.72 & 0.74 & 0.56 & d \\
13 30 51.82 & -31 04 40.3 & -20.5 & \phn1.56 & 4.63 & 2.24 & 2.45 & 1.72 & 2.23 & a \\
13 30 52.20 & -31 46 07.0 & -20.3 & \phn0.69 & 4.28 & 2.04 & 1.70 & 0.84 & 0.48 & \nodata \\
13 31 00.75 & -31 44 53.9 & -20.8 & \phn0.97 & 4.37 & 2.10 & 1.60 & 0.88 & 0.42 & e \\
13 31 16.90 & -31 35 38.4 & -20.7 & \phn0.81 & 4.54 & 2.15 & 1.44 & 0.95 & 0.68 & \nodata \\
13 31 19.47 & -31 13 44.4 & -21.7 & \phn1.56 & 4.76 & 2.30 & 1.92 & 1.49 & 1.74 & \nodata \\
13 31 20.22 & -31 44 34.6 & -20.7 & \phn0.98 & 4.57 & 2.29 & 1.40 & 1.06 & 0.27 & \nodata \\
13 31 37.79 & -31 33 12.0 & -21.4 & \phn1.20 & 4.77 & 2.36 & 1.25 & 1.18 & 0.77 & f \\
13 31 54.87 & -31 54 04.1 & -20.6 & \phn0.80 & 4.97 & 2.80 & 1.22 & 1.60 & 0.33 & \nodata \\
13 32 01.52 & -31 13 58.0 & -21.6 & \phn0.73 & 5.16 & 2.71 & 1.62 & 1.80 & 1.74 & a \\
13 32 03.17 & -31 46 48.6 & -22.4 & \phn1.77 & 5.03 & 2.75 & 0.98 & 1.52 & 0.30 & \nodata \\
13 32 30.07 & -31 03 15.8 & -20.2 & \phn1.26 & 5.63 & 3.19 & 1.95 & 2.38 & 2.32 & \nodata \\
13 32 35.44 & -31 31 45.5 & -20.2 & \phn0.74 & 5.38 & 2.96 & 0.73 & 1.79 & 1.05 & \nodata \\
13 32 46.48 & -31 37 03.3 & {\it -21.0} & \phn0.75 & 5.48 & 3.09 & 0.51 & 1.89 & 0.94 & \nodata \\
13 32 46.75 & -31 57 59.2 & -22.2 & \phn0.91 & 5.54 & 3.37 & 0.99 & 2.17 & 0.87 & g \\
13 33 04.16 & -31 39 04.0 & -20.1 & \phn0.85 & 5.66 & 3.29 & 0.30 & 2.08 & 1.05 & \nodata \\
13 33 31.63 & -31 40 58.6 & {\it -21.0} & 42.5\phn & 5.95 & 3.59 & 0.03 & 2.37 & 1.28 & \nodata \\
13 33 40.7\phn  & -31 19 57.\phn  & -20.6 & \phn2.57 & 6.13 & 3.68 & 1.02 & 2.59 & 1.94 & a \\
13 34 39.94 & -31 32 54.5 & -20.5 & \phn2.21 & 6.68 & 4.28 & 0.81 & 3.09 & 2.09 & \nodata \\
13 35 03.06 & -31 39 18.6 & -22.4 & \phn6.38 & 6.92 & 4.53 & 0.96 & 3.33 & 2.23 & \nodata \\
13 35 12.61 & -31 33 45.0 & -21.1 & \phn0.96 & 7.03 & 4.62 & 1.11 & 3.43 & 2.40 & \nodata \\
13 35 42.55 & -31 53 54.6 & -21.9 & \phn5.93 & 7.35 & 5.06 & 1.53 & 3.83 & 2.60 & \nodata \\
13 35 56.86 & -32 01 50.0 & -20.4 & \phn0.82 & 7.56 & 5.33 & 1.86 & 4.09 & 2.82 & h \\
13 36 18.04 & -31 46 45.3 & -22.1 & \phn1.06 & 7.70 & 5.36 & 1.78 & 4.14 & 2.97 & \nodata \\
13 36 18.70 & -31 48 31.2 & -21.4 & \phn1.14 & 7.71 & 5.38 & 1.80 & 4.16 & 2.97 & i \\
\cutinhead{Additional Cluster Radio Galaxies:}
13 19 07.85 & -31 40 10.8 & {\it -21.0} & \phn0.28 & 3.14 & 5.57 & 9.09 & 6.73 & 7.89 & \nodata \\
13 20 54.63 & -31 44 33.4 & -21.1 & \phn3.27 & 2.03 & 4.48 & 7.97 & 5.61 & 6.75 & \nodata \\
13 20 55.59 & -31 43 59.3 & -21.4 & \phn2.74 & 2.02 & 4.47 & 7.96 & 5.60 & 6.74 & \nodata \\
13 21 45.02 & -31 03 00.6 & -21.8 & \phn3.21 & 2.35 & 4.13 & 7.71 & 5.37 & 6.66 & \nodata \\
13 21 52.65 & -31 45 54.4 & -21.6 & \phn0.59 & 1.44 & 3.89 & 7.36 & 5.01 & 6.14 & \nodata \\
13 22 49.71 & -32 26 11.6 & -22.1 & \phn1.31 & 2.44 & 4.24 & 7.08 & 5.00 & 5.80 & \nodata \\
13 24 39.14 & -32 06 33.4 & -21.5 & \phn0.24 & 1.38 & 2.75 & 5.73 & 3.55 & 4.46 & \nodata \\
13 24 50.44 & -31 33 37.8 & -20.5 & \phn0.48 & 0.55 & 1.95 & 5.50 & 3.13 & 4.34 & \nodata \\
13 25 50.32 & -32 08 54.2 & -22.4 & \phn0.26 & 1.82 & 2.35 & 5.06 & 2.96 & 3.77 & \nodata \\
13 26 25.80 & -31 27 20.9 & -21.4 & \phn0.32 & 1.59 & 0.95 & 4.54 & 2.17 & 3.44 & \nodata \\
13 26 35.69 & -30 59 36.4 & -20.1 & \phn0.22 & 2.54 & 1.70 & 4.85 & 2.73 & 4.01 & \nodata \\
13 26 49.94 & -30 37 56.9 & -20.3 & \phn0.79 & 3.51 & 2.64 & 5.28 & 3.46 & 4.64 & \nodata \\
13 26 52.95 & -30 00 28.6 & -21.7 & \phn0.16 & 5.22 & 4.46 & 6.53 & 5.10 & 6.16 & \nodata \\
13 26 54.01 & -31 18 39.5 & -19.8 & \phn0.19 & 2.05 & 0.84 & 4.34 & 2.03 & 3.33 & \nodata \\
13 26 57.51 & -32 11 52.3 & -20.8 & \phn0.32 & 2.40 & 2.18 & 4.41 & 2.49 & 3.13 & \nodata \\
13 27 03.35 & -31 13 31.5 & -21.4 & \phn0.47 & 2.28 & 0.96 & 4.32 & 2.07 & 3.37 & \nodata \\
13 27 29.66 & -30 33 31.8 & -21.1 & \phn1.74 & 3.92 & 2.78 & 5.08 & 3.44 & 4.55 & \nodata \\
13 27 34.96 & -31 32 59.2 & -21.8 & \phn0.33 & 2.22 & 0.27 & 3.78 & 1.40 & 2.66 & \nodata \\
13 27 45.50 & -31 47 51.0 & -22.5 & \phn0.20 & 2.34 & 0.91 & 3.66 & 1.40 & 2.42 & \nodata \\
13 27 51.48 & -31 14 45.1 & -19.6 & \phn0.64 & 2.68 & 0.73 & 3.82 & 1.63 & 2.91 & \nodata \\
13 28 03.55 & -30 40 50.9 & -20.5 & \phn0.60 & 3.85 & 2.41 & 4.57 & 2.96 & 4.05 & \nodata \\
13 28 10.37 & -31 20 57.4 & -20.9 & \phn0.38 & 2.74 & 0.46 & 3.53 & 1.28 & 2.57 & \nodata \\
13 28 33.79 & -31 26 28.3 & -20.7 & \phn0.35 & 2.90 & 0.44 & 3.22 & 0.92 & 2.22 & \nodata \\
13 28 35.78 & -31 30 54.3 & -20.1 & \phn0.27 & 2.88 & 0.44 & 3.16 & 0.80 & 2.10 & \nodata \\
13 28 38.72 & -31 20 49.3 & -22.1 & \phn0.25 & 3.03 & 0.63 & 3.24 & 1.06 & 2.33 & \nodata \\
13 28 39.26 & -31 39 41.1 & -19.9 & \phn0.40 & 2.87 & 0.69 & 3.08 & 0.73 & 1.91 & \nodata \\
13 28 54.30 & -31 55 11.1 & -20.8 & \phn0.29 & 3.12 & 1.41 & 3.00 & 1.08 & 1.73 & \nodata \\
13 28 59.29 & -31 44 15.7 & -20.2 & \phn0.65 & 3.09 & 1.00 & 2.87 & 0.63 & 1.66 & \nodata \\
13 29 04.33 & -31 47 52.4 & -20.5 & \phn0.28 & 3.16 & 1.17 & 2.84 & 0.72 & 1.60 & \nodata \\
13 29 12.29 & -31 18 39.9 & {\it -21.0} & \phn0.23 & 3.40 & 0.98 & 2.95 & 0.95 & 2.13 & \nodata \\
13 29 14.98 & -31 16 44.9 & -21.5 & \phn0.44 & 3.45 & 1.06 & 2.96 & 1.03 & 2.18 & \nodata \\
13 29 17.00 & -32 02 33.9 & -21.9 & \phn0.47 & 3.45 & 1.85 & 2.89 & 1.33 & 1.61 & \nodata \\
13 29 20.43 & -30 46 37.4 & -20.8 & \phn0.67 & 4.25 & 2.31 & 3.77 & 2.48 & 3.40 & \nodata \\
13 29 31.44 & -31 37 29.0 & -20.3 & \phn0.36 & 3.43 & 1.09 & 2.54 & 0.17 & 1.43 & \nodata \\
13 29 54.20 & -32 00 59.3 & -21.0 & \phn0.42 & 3.80 & 2.00 & 2.50 & 1.21 & 1.23 & \nodata \\
13 30 05.20 & -31 13 53.4 & -20.2 & \phn0.39 & 4.00 & 1.58 & 2.55 & 1.13 & 1.97 & \nodata \\
13 30 24.83 & -30 41 37.1 & -21.4 & \phn0.55 & 4.95 & 2.86 & 3.52 & 2.74 & 3.41 & \nodata \\
13 30 32.80 & -31 34 33.8 & -22.4 & \phn0.61 & 4.08 & 1.68 & 1.91 & 0.49 & 0.97 & \nodata \\
13 30 40.65 & -31 42 32.1 & -19.3 & \phn0.22 & 4.15 & 1.86 & 1.80 & 0.64 & 0.66 & \nodata \\
13 30 40.74 & -31 51 56.0 & -20.7 & \phn0.18 & 4.19 & 2.07 & 1.89 & 0.95 & 0.60 & \nodata \\
13 30 41.47 & -31 34 17.3 & -21.2 & \phn0.64 & 4.17 & 1.77 & 1.82 & 0.58 & 0.92 & \nodata \\
13 30 47.73 & -31 51 06.5 & -21.4 & \phn0.23 & 4.26 & 2.11 & 1.80 & 0.96 & 0.52 & \nodata \\
13 30 51.04 & -31 44 17.9 & -21.5 & \phn0.23 & 4.26 & 1.99 & 1.70 & 0.78 & 0.52 & \nodata \\
13 30 52.70 & -31 27 13.2 & -20.6 & \phn0.39 & 4.33 & 1.88 & 1.80 & 0.83 & 1.16 & \nodata \\
13 30 53.83 & -31 30 21.7 & -21.0 & \phn0.21 & 4.32 & 1.89 & 1.74 & 0.77 & 1.01 & \nodata \\
13 30 56.16 & -31 33 42.9 & -19.9 & \phn0.23 & 4.33 & 1.92 & 1.67 & 0.74 & 0.85 & \nodata \\
13 31 01.99 & -31 40 36.3 & -19.8 & \phn0.30 & 4.38 & 2.05 & 1.58 & 0.82 & 0.54 & \nodata \\
13 31 17.43 & -31 24 34.7 & -20.7 & \phn0.58 & 4.61 & 2.15 & 1.62 & 1.12 & 1.21 & \nodata \\
13 31 22.42 & -31 22 10.4 & -19.6 & \phn0.21 & 4.69 & 2.22 & 1.64 & 1.23 & 1.32 & \nodata \\
13 31 27.58 & -31 49 14.7 & -22.1 & \phn0.65 & 4.66 & 2.44 & 1.38 & 1.23 & 0.09 & \nodata \\
13 31 59.81 & -31 49 22.2 & -19.8 & \phn0.37 & 5.00 & 2.76 & 1.06 & 1.53 & 0.25 & \nodata \\
13 32 38.27 & -31 01 41.3 & -20.0 & \phn0.55 & 5.74 & 3.31 & 2.00 & 2.50 & 2.42 & \nodata \\
13 32 42.95 & -31 55 13.7 & -20.5 & \phn0.14 & 5.48 & 3.28 & 0.90 & 2.07 & 0.77 & \nodata \\
13 32 43.89 & -31 47 01.3 & -20.7 & \phn0.28 & 5.46 & 3.16 & 0.60 & 1.93 & 0.72 & \nodata \\
13 33 01.00 & -31 13 29.3 & -20.4 & \phn0.50 & 5.80 & 3.33 & 1.37 & 2.35 & 1.96 & \nodata \\
13 33 23.53 & -31 24 23.3 & -20.9 & \phn0.16 & 5.93 & 3.48 & 0.80 & 2.36 & 1.66 & \nodata \\
13 33 34.76 & -31 40 20.5 & -22.3 & \phn0.16 & 5.99 & 3.62 & 0.03 & 2.41 & 1.32 & \nodata \\
13 33 50.43 & -31 28 38.0 & -20.9 & \phn0.25 & 6.19 & 3.75 & 0.61 & 2.60 & 1.73 & \nodata \\
13 33 58.73 & -31 58 16.1 & -20.4 & \phn0.49 & 6.29 & 4.07 & 0.93 & 2.85 & 1.57 & \nodata \\
13 34 31.48 & -32 05 09.0 & -20.8 & \phn0.41 & 6.67 & 4.52 & 1.38 & 3.30 & 2.01 & \nodata \\
13 34 36.19 & -31 41 06.7 & -22.6 & \phn0.61 & 6.63 & 4.26 & 0.68 & 3.05 & 1.93 & \nodata \\
13 34 40.80 & -31 26 43.6 & -21.3 & \phn0.65 & 6.73 & 4.29 & 0.99 & 3.14 & 2.23 & \nodata \\
13 34 43.00 & -32 22 14.0 & -21.4 & \phn0.46 & 6.98 & 5.00 & 2.20 & 3.83 & 2.56 & \nodata \\
13 35 29.53 & -30 52 56.1 & -21.5 & \phn1.01 & 7.61 & 5.16 & 2.66 & 4.23 & 3.71 & \nodata \\
\enddata

\tablenotetext{a}{No velocity available, but color does not rule out membership.}

\tablenotetext{b}{The velocity of this galaxy, $cz = 12441$ \kms, is well within the accepted range for the supercluster. However, it falls outside the range allowed for SC1327-312 for the purpose of calculation of the radio galaxy fraction.}

\tablenotetext{c}{The source is a member of a galaxy pair, with no velocity available for the galaxy nearest the radio emission. The other galaxy, at 13:28:57.9 -31:51:36, has a cluster velocity of $cz=16532$ \kms.}

\tablenotetext{d}{Optical magnitude possibly corrupted by foreground star. See Figure \ref{fig-133048} and discussion in Section \ref{sec-discuss}.}

\tablenotetext{e}{The velocity of this galaxy, $cz = 11976$ \kms, is within the accepted range for the supercluster. However, it falls outside the ranges allowed for SC1327-312 and SC1329-313 for the purpose of calculation of the radio galaxy fractions.}

\tablenotetext{f}{The source is a member of a galaxy pair, with no velocity available for the galaxy nearest the radio emission. The other galaxy, at 13:31:37.56 -31:33:06.5, has a cluster velocity of $cz=14194$ \kms.}

\tablenotetext{g}{The velocity of this galaxy, $cz = 11106$ \kms, is within the accepted range for the supercluster. However, it falls outside the ranges allowed for A3562 and SC1329-313 for the purpose of calculation of the radio galaxy fractions.}

\tablenotetext{h}{No velocity available, but color ($B - R = 2.33$) strongly suggests this is a background source.}

\tablenotetext{i}{The velocity of this galaxy, $cz = 11692$ \kms, is within the accepted range for the supercluster. However, it falls outside the range allowed for A3562 for the purpose of calculation of the radio galaxy fraction.}

\tablecomments{Italicized values within the absolute magnitude column indicate higher absolute magnitudes (i.e., fainter galaxies) which have been rounded to the presented figure. All reported radio luminosities have been calculated using an assumed spectral index of 0.7. For unresolved sources, they are based on the fitted peak flux density (see text).}

\end{deluxetable}

\begin{deluxetable}{l r r r r@{ - }r}
\tablecolumns{6}
\tablecaption{Radio Galaxy Fractions, $M_R \leq -22$\label{tbl-brightfracs}}
\tablewidth{0pt}
\tablehead{
\colhead{Sample} & \colhead{$N_{RG}$} & \colhead{$N_{tot}$} & \colhead{Sig} & 
\multicolumn{2}{c}{Range}
}
\startdata
A3556      & 7 &  12.7 & {\sl 98.7} & 92.3 & 99.9 \\
A3558      & 8 &  21.7 &       82.0 & 63.6 & 93.7 \\
A3562      & 5 &  11.7 &       85.7 & 59.9 & 98.1 \\
SC1327-312 & 8 &  19.7 &       89.9 & 75.5 & 97.4 \\
SC1329-313 & 3 &  12.7 &        0.1 & 34.0 & 43.8 \\
Comparison & 40 & 169.4 & \nodata & \nodata & \nodata \\
\enddata

\tablecomments{$N_{RG}$ is the number of radio-detected galaxies and $N_{tot}$ is the background-corrected total number of galaxies. Only radio galaxies with velocity measurements confirming cluster membership entered the analysis for this optical magnitude range. ``Sig'' indicates the significance with which a given radio galaxy fraction differs from the expectations of the collective comparison sample (i.e., higher numbers indicate unusual radio galaxy fractions). Values greater than $95\%$ have been italicized. ``Range'' represents the significance levels determined after changing the background correction by $\pm100\%$. For SC1329-313, the effect of changing the background alters the direction of the discrepancy from the comparison sample. Subtracting a larger background makes SC1329-313 have an excess of radio galaxies at 34$\%$ significance, while not subtracting any background make this cluster appear to have a deficit of radio galaxies at 43.8$\%$ significance.}
\end{deluxetable}

\begin{deluxetable}{l r r r r@{ - }r}
\tablecolumns{6}
\tablecaption{Radio Galaxy Fractions, $-22 < M_R \leq -21$\label{tbl-intfracs}}
\tablewidth{0pt}
\tablehead{
\colhead{Sample} & \colhead{$N_{RG}$} & \colhead{$N_{tot}$} & \colhead{Sig} & 
\multicolumn{2}{c}{Range}
}
\startdata
A3556      & 0 &  34.3 & {\sl 98.5} & 96.9 & 99.2 \\
A3558      & 3 &  87.3 & {\sl 99.7} & 99.4 & 99.8 \\
A3562      & 3 &  58.3 & {\sl 96.0} & 92.8 & 97.8 \\
~~extended  & 5 &       &       81.3 & 69.5 & 88.8 \\
SC1327-312 & 4 & 108.3 & {\sl 99.9} & 99.7 & 99.9 \\
~~extended   & 6 &       & {\sl 99.2} & 98.6 & 99.5 \\
SC1329-313 & 4 &  83.3 & {\sl 98.8} & 98.0 & 99.3 \\
~~extended   & 6 &       &       94.3 & 90.7 & 96.6 \\
Comparison & 87 & 582.4 & \nodata & \nodata & \nodata \\
\enddata

\tablecomments{See comments to Table \ref{tbl-brightfracs}. Rows labelled ``extended'' include galaxies whose photometry indicates likely cluster membership on the basis of optical color, or whose membership is assumed as they are members of a galaxy pair in which the companion has a cluster velocity.}
\end{deluxetable}

\begin{deluxetable}{l r r r r@{ - }r}
\tablecolumns{6}
\tablecaption{Radio Galaxy Fractions, $-21 < M_R \leq -20$\label{tbl-faintfracs}}
\tablewidth{0pt}
\tablehead{
\colhead{Sample} & \colhead{$N_{RG}$} & \colhead{$N_{tot}$} & \colhead{Sig} & 
\multicolumn{2}{c}{Range}
}
\startdata
A3556      &  1 &   81.2 &          29.7 &  1.0 & 49.9 \\
~~extended   & 2 &        &          32.6 &  2.5 & 67.5 \\
A3558      &  3 &  180.2 &          10.9 & 10.0 & 28.4 \\
~~extended   & 4 &        &          29.3 &  9.6 & 50.1 \\
A3562      & 11 &  113.2 & {\sl $>$99.9} & $>$99.9 & $>$99.9 \\
~~extended   & 12 &        & {\sl $>$99.9} & $>$99.9 & $>$99.9 \\
SC1327-312 &  9 &  196.2 &    {\sl 98.5} & 96.2 & 99.6 \\
~~extended   & 11 &        &    {\sl 99.9} & $>$99.9 & $>$99.9 \\
SC1329-313 &  9 &  142.2 &    {\sl 99.9} & $>$99.9 & $>$99.9 \\
~~extended   & 11 &        & {\sl $>$99.9} & $>$99.9 & $>$99.9 \\
Comparison & 20 & 1104.6 & \nodata & \nodata & \nodata \\
\enddata

\tablecomments{See comments to Table \ref{tbl-brightfracs}. Rows labelled ``extended'' include galaxies whose photometry indicates likely cluster membership on the basis of optical color, or whose membership is assumed as they are members of a galaxy pair in which the companion has a cluster velocity.}
\end{deluxetable}

\clearpage

\begin{figure}
\figurenum{1}
\caption{Areal coverage of the optical and radio data. The straight solid lines indicate the edges of the individual optical images, while the grey region signifies the coverage of the radio mosaic. The locations of the centers of the three Abell clusters and two X-ray identified clusters are marked by stars (from right to left, A3556, A3558, SC1327-312, SC1329-313, and A3562), with the large circles identifying a radial distance of 2 Mpc for each of the three Abell clusters.\label{fig-foot}}
\end{figure}

\begin{figure}
\figurenum{2}
\caption{J132145-310300, a cluster radio galaxy with $L_{1.4} = 3.2 \times 10^{22}$ W Hz$^{-1}$. In this and all subsequent overlay plots, radio contours are plotted at 2, 3, 5, 8, 13, 21, 34, 55, 89, 144, 233, and 377 times the local noise level, which for this image is 121 $\mu$Jy beam$^{-1}$.\label{fig-132145}}
\end{figure}

\begin{figure}
\figurenum{3}
\caption{J132206-314616, a cluster radio galaxy with $L_{1.4} = 3.0 \times 10^{22}$ W Hz$^{-1}$. The base level for this image is 70 $\mu$Jy beam$^{-1}$.\label{fig-132206}}
\end{figure}

\begin{figure}
\figurenum{4}
\caption{J132210-312805, a likely background radio galaxy. The base level for this image is 71 $\mu$Jy beam$^{-1}$.\label{fig-132210}}
\end{figure}

\begin{figure}
\figurenum{5}
\caption{J132357-313845, a cluster radio galaxy with $L_{1.4} = 2.3 \times 10^{23}$ W Hz$^{-1}$ and discussed in detail in \citet{venturi1998}. The base level for this image is 82 $\mu$Jy beam$^{-1}$.\label{fig-132357}}
\end{figure}

\begin{figure}
\figurenum{6}
\caption{J132802-314521, a cluster radio galaxy with $L_{1.4} = 1.3 \times 10^{23}$ W Hz$^{-1}$. The base level for this image is 81 $\mu$Jy beam$^{-1}$.\label{fig-132802}}
\end{figure}

\begin{figure}
\figurenum{7}
\caption{J132950-312258, a background radio galaxy at $z=0.196$ with $L_{1.4} = 9.7 \times 10^{23}$ W Hz$^{-1}$. The base level for this image is 114 $\mu$Jy beam$^{-1}$.\label{fig-132950}}
\end{figure}

\begin{figure}
\figurenum{8}
\caption{J133048-314325 (top galaxy), a cluster radio galaxy whose morphology is similar to that of a WAT but with $L_{1.4}$ of only $3.1 \times 10^{22}$ W Hz$^{-1}$. The ``pinching'' of the contours along declination -31:45:00 is caused by the stitching together of individual facets of the radio mosaic. The base level for the image is 92 $\mu$Jy beam$^{-1}$.\label{fig-133048}}
\end{figure}

\begin{figure}
\figurenum{9}
\caption{J133331-314058, a cluster radio galaxy with $L_{1.4} = 4.3 \times 10^{23}$ W Hz$^{-1}$. Note that the ``Y'' shape is caused by the ({\it u,v}) coverage and is not intrinsic to the source. The base level for this image is 92 $\mu$Jy beam$^{-1}$. See \citet{venturi2003} for additional details, including observations at multiple frequencies. This reference also discusses the A3562 halo, which lies to the west of the powerful radio galaxy.\label{fig-133331}}
\end{figure}

\begin{figure}
\figurenum{10}
\caption{J133542-315354, a cluster radio galaxy with $L_{1.4} = 6.0 \times 10^{22}$ W Hz$^{-1}$. The base level for this image is 77 $\mu$Jy beam$^{-1}$.\label{fig-133542}}
\end{figure}

\begin{figure}
\figurenum{11}
\caption{Color magnitude diagrams for each cluster, along with fitted red sequence. The plots include all galaxies within 2 Mpc of the respective cluster centers, with the exception of the plot at top left which includes all the optical data. Error bars have been left off for clarity; a representative point including error bars is included at the bottom left of each plot.\label{fig-redseq}}
\end{figure}

\begin{figure}
\figurenum{12}
\caption{$B - R$ colors for the detected radio galaxies, plotted in four $R$ magnitude ranges. Galaxies with NED velocities are shaded, with cluster members solid and foreground/background objects hatched.\label{fig-colors}}
\end{figure}

\begin{figure}
\figurenum{13}
\caption{Radio galaxy fractions evaluated over 0.5 Mpc radius regions. For this plot, only galaxies with $M_R \leq -20$ are included, and to be considered a radio detection a flux of $\geq$ 1.64 mJy is required (i.e., consistent with prior studies). Black represents fractions of 40$\%$ or greater, with white representing fractions of zero. Contours of galaxy surface density are included to indicate where high fractions result from low number statistics. These contours are plotted in green at 5, 10, 20, 30, 40, 50, 60, 70, and 80 galaxies Mpc$^{-2}$. The centers of the five clusters are marked with blue stars, with 2 Mpc radii circles around the three Abell clusters.\label{fig-fracN}}
\end{figure}

\begin{figure}
\figurenum{14}
\caption{Same as for Figure \ref{fig-fracN}, but for radio detections with fluxes between 0.5 mJy and 1.64 mJy. Note that the optical and radio limits covered in this figure include 63 sources from Table \ref{tbl-radgals}. Of these, 58 are included in this plot as potential cluster radio galaxies (43 with velocity information, 15 without but accepted on the basis of color). Four of the five radio galaxies not included were rejected on the basis of velocity information, with the fifth being too red to be a cluster member.\label{fig-fracT}}
\end{figure}

\begin{figure}
\figurenum{15}
\caption{Same as for Figure \ref{fig-fracN}, but for galaxies with $-20 < M_R \leq -19$. To be a radio detection, a flux greater than 0.5 mJy is required (note that all of these sources have fluxes below the 1.64 mJy limit used in Figure \ref{fig-fracN}). The optical and radio limits covered in this figure include 53 sources from Table \ref{tbl-radgals}. Of these 53 galaxies, 18 are not included in this plot as they were deemed background galaxies on the basis of velocity or color information (2 and 16, respectively). Of the 35 cluster radio galaxies included in the plot, most (28) lacked public velocity measurements but were included as possible cluster members on the basis of their colors.\label{fig-fracP}}
\end{figure}

\begin{figure}
\figurenum{16}
\caption{Integral radio luminosity functions for galaxies with $M_R \leq -20$, expressed as the fraction of cluster galaxies with radio luminosity of a given flux or greater. The Comparison Sample RLF (in black) was determined from the 18 clusters of the comparison sample using radio galaxies with projected separations of under 2 Mpc from their respective cluster centers. Similarly, the Shapley supercluster RLF (red) is based on all radio galaxies within 2 Mpc of any of the five cluster centers while that for A3558 (blue) is for just that cluster. Points for the supercluster and A3558 have been offset slightly for clarity. Inclusion of the radio galaxies without spectroscopic confirmation (i.e., those in pairs and those with colors consistent with cluster membership) increases the Shapley RLF to about the level of the upper error bars. Note that the radio galaxy fractions analysis is for sources with luminosities over about $7 \times 10^{21}$. The scaled RLF of \citet{ledlow1996} is provided for comparison (green triangles, see text).\label{fig-rlf}}
\end{figure}

\end{document}